\documentclass[11pt]{article}
\usepackage{amsmath,amssymb}
\def\baselinestretch{1.2}
\usepackage{float}
\usepackage{graphicx}
\usepackage{wrapfig}
\usepackage{slashed}
\usepackage{hyperref}
\usepackage{cite}
\usepackage{longtable}
\usepackage{xcolor}
\usepackage{geometry}
\def\beq{\begin{eqnarray}}
\def\eeq{\end{eqnarray}}

\def\la{\langle }
\def\ra{\rangle }

\newcommand{\Tr}{\,\mathrm{Tr}\,}   
\newcommand{\tr}{\,\mathrm{tr}\,}

\newcommand{\be}{\begin{equation}}
\newcommand{\ee}{\end{equation}}
\newcommand{\bea}{\begin{eqnarray}}
\newcommand{\eea}{\end{eqnarray}}
\newcommand{\bg}{\begin{gather}}

\newcommand{\bseq}{\begin{subequations}}
\newcommand{\eseq}{\end{subequations}}

\renewcommand{\ln}{\mathop{\rm ln}\nolimits}

\def\be{\begin{eqnarray}}
\def\ee{\end{eqnarray}}
\def\lb{\label}
\newcommand{\Kh}{\hat{K}}
\newcommand{\nb}{\bar{\nabla}}

\newcommand{\CM}{{\mathcal{M}}}

\newcommand{\CA}{{\mathcal{A}}}

\newcommand{\p}{\partial}

\usepackage{booktabs}
\begin{document}

\title{\textbf{Fermions, boundaries and conformal and chiral \\
anomalies in   $d=3,\ 4$ and $5$ dimensions}}

\vspace{2cm}
\author{ \textbf{  Amin Faraji Astaneh$^{a}$ and  Sergey N. Solodukhin$^b$ }} 
\date{}
\maketitle
\begin{center}
\hspace{-0mm}
\emph{$^a$ 
Research Center for High Energy Physics and\\
Department of Physics, Sharif University of Technology,\\
P.O.Box 11155-9161, Tehran, Iran}
 \end{center}
\begin{center}
  \hspace{-0mm}
   \emph{ $^b$  Institut Denis Poisson UMR 7013,
  Universit\'e de Tours,}\\
  \emph{Parc de Grandmont, 37200 Tours, France} \\
\end{center}

\begin{abstract}
\noindent {

In the presence of boundaries, the quantum anomalies acquire additional boundary terms.  In odd dimensions the integrated conformal anomaly, for which  the bulk
contribution is known to be absent, is non-trivial due to the   boundary terms. These terms became a subject of active study in the recent years. 
In the present paper we continue our previous study \cite{Solodukhin:2015eca}, \cite{FarajiAstaneh:2021foi} and compute explicitly the anomaly for fermions in dimensions $d=3, \ 4 \ $ and $5$. The calculation in dimension $5$ is new. It contains both contributions of the gravitational field and the gauge fields to the anomaly.
In dimensions $d=3$ and $4$ we reproduce and clarify the derivation of the results available in the literature. Imposing the conformal invariant mixed boundary conditions for fermions in odd dimension $d$, we
particularly pay attention to the necessity  of choosing  the doubling representation for gamma matrices.
In this representation there exists a possibility to define chirality and thus address the question of the chiral anomaly. The anomaly is  entirely due to  terms  defined on the boundary.
They are calculated in the present paper in dimensions $d=3$ and $5$ due to both gravitational and  gauge fields. To complete the picture we re-evaluate the chiral anomaly in $d=4$ dimensions and find a new boundary term that is supplementary to the well-known Pontryagin term.
}

\end{abstract}

\vskip 2 cm
\noindent
\rule{7.7 cm}{.5 pt}\\
\noindent 
\noindent
\noindent ~~~ {\footnotesize e-mails:\ faraji@sharif.ir ,  Sergey.Solodukhin@lmpt.univ-tours.fr}

\newpage

\section{Introduction}

The role of quantum anomalies in the modern theoretical constructions becomes increasingly important.
Conformal anomalies first discovered by Capper and Duff almost 50 years ago
\cite{Capper:1974ic}  by now serve as excellent example of very rich and mutually useful  interplay between the differential geometry, quantum fundamental physics and
applications. In the recent years a new aspect of conformal anomaly came into play. The presence of boundaries changes rather dramatically what 
we used to think about the anomaly. Indeed, the local geometric invariants from which the anomaly can be constructed have necessarily even dimensionality.
So that the anomaly is conventionally absent in space-time of odd dimension $d$ since no invariant of appropriate odd dimensionality exits.
This is no more true in the presence of boundaries. A geometric quantity, extrinsic curvature,  characterizes how the boundary is embedded into spacetime
and it has dimension one. This allows to construct new invariants of both odd and even dimensionality on the boundary of spacetime.
As a consequence, the conformal anomaly or, better to say,  the integrated conformal anomaly can be now non-trivial even if the dimension of spacetime is odd.
If dimension is even there, additionally to the bulk terms,  appear boundary terms with increasingly reach structure as the dimension $d$ grows.
An earlier paper in this direction is  \cite{Dowker:1989ue}.
The complete structure, or the building blocks from which one can construct the boundary anomaly terms is not yet fully understood for large values of $d$.
In dimensions $d=3$ and $d=4$ the situation is by now  quite clear after the works  \cite{Solodukhin:2015eca}, \cite{Fursaev:2015wpa}, \cite{Herzog:2015ioa}.
In \cite{Fursaev:2015wpa} the values of the boundary conformal charges in dimension $d=4$ have been computed for free conformal fields: scalar fields, Dirac fermions
and gauge fields.
In dimension $d=5$ a recent progress has been reached after identifying a complete set of boundary conformal invariants in this dimension in  \cite{FarajiAstaneh:2021foi}.  The respective conformal charges for a conformal scalar field in $d=5$ have been computed in \cite{FarajiAstaneh:2021foi}.
The further developments in this direction include \cite{Fursaev:2016inw} -\cite{Juhl:2023mts}.

The primary goal of  the present paper is to build on our previous work \cite{FarajiAstaneh:2021foi} and compute the boundary conformal charges  for the fermions
in dimension $d=5$. Let us briefly discuss a difficulty that looks technical but whose resolution leads to interesting consequences.
Considering  the Dirac fermions in space-time with boundaries one encounters the following problem. 
One imposes the boundary conditions on the fermion fields of the mixed type: a half of components of the fermion satisfy the Dirichlet condition while the other half
a conformal Robin type condition. Thus, one needs to define two projectors $\Pi_+=\frac{1}{2}(1+\chi)$ and $\Pi_-=\frac{1}{2}(1-\chi)$ such that $\Pi_++\Pi_-=1$. 
The condition on matrix $\chi$ is that it has to anti-commute with $\gamma^n=n_k \gamma^k$, where $n^k$ is normal vector to the boundary, and commute with
all other gamma matrices $\gamma^a\, , \ a=1, \dots , d-1$ projected along the boundary. In even dimension $d$, this matrix can be easily constructed, $\chi=i \gamma^* \gamma^n$.
Here $\gamma^*$ is the chirality matrix, it anti-commutes with all gamma matrices. In odd dimension $d$, provided one uses  the usual $2^{\frac{d-1}{2}}\times 2^{\frac{d-1}{2}}$
representation for gamma-matrices, such matrix $\gamma^*$ does not exist. This forces us to use other representation for gamma matrices of dimension
$2^{\frac{d+1}{2}}\times 2^{\frac{d+1}{2}}$  that is obtained in the so called   doubling procedure.
 For earlier discussions
of this representation  see \cite{doubling}, \cite{d=3 fermions}. 
In this representation it is known that two chiral matrices exist, what we call below $\Gamma^*_1$ and $\Gamma^*_2$. The notion of chirality is thus naturally defined. That is why in the present paper we also compute the chiral anomaly in dimensions $d=3\, , \ 4$ and $5$.

In  dimension $d=4$ the above mentioned problem does not arise and one uses the standard representation for gamma matrices. The respective chiral anomaly is due to the Pontryagin term, as is well known. However, the careful analysis presented below reveals a new boundary term in the chiral anomaly that is supplementary to the Pontryagin term.

In order to make our consideration  general we also include the coupling of the fermion field to a background gauge field and compute the contribution of the gauge field to the boundary terms in the conformal and chiral anomalies. Our findings are summarised in Table \ref{table}.

It should be noted that the key tool in our computations is the heat kernel method. Thus, we heavily  rely on the available results for the heat kernel coefficients for manifolds with boundaries
given in \cite{Vassilevich:2003xt} and \cite{Branson:1999jz}.

\begin{table}\label{table}
\begin{center}
\renewcommand{\baselinestretch}{2}
\medskip
\caption{ Contributions to boundary terms in conformal and chiral anomaly}
\bigskip
\begin{tabular}{| l | cc | cc | cc| }
\hline
\bf{Dimensions}  & \multicolumn{2}{c|}{$\mathbf{d=3}$} & \multicolumn{2}{c|}{$\mathbf{d=4}$} & \multicolumn{2}{c|}{$\mathbf{d=5}$} \\
\hline
Type of anomaly     & Conformal   & Chiral  & Conformal  & Chiral & Conformal & Chiral \\
\hline
Boundary terms due to gravitational field & Yes & No & Yes & Yes & Yes & Yes\\
\hline
Boundary terms due to gauge field & No & Yes & No & No & Yes & Yes\\
\hline
\end{tabular}
\renewcommand{\baselinestretch}{1}
\end{center}
\end{table}

\section{The basics}
\subsection{Dirac operator in curved spacetime}
The theory lives on manifold $M$ covered by coordinates $x^i$ and with the metric $g_{ij}$. This manifold has a boundary  $\p M$, the study of its contribution to the conformal anomaly is the main topic of this note.  We also introduce a basis of orthonormal tangent vectors, the {\it{n-bein}}s $e^p_i(x)$ at each point on $M$ so that
\be
g_{ij}=\eta_{pq}\, e^p_i e^q_i\, .
\ee
We consider a Dirac theory describing a spinor $\psi(x)$ on this manifold. The action reads
\be
S=\int d^dx\sqrt{-g}\, \bar{\psi}\,  i\gamma^k\hat{\nabla}_k\,\psi \ , 
\ee
where $\gamma^k=e^k_p\gamma^p$ and $\gamma^p$ are the Dirac matrices satisfying the Clifford algebra,
\be
&&\gamma^p\gamma^q+\gamma^q\gamma^p=2\eta^{qp} \, , \  q,p=0, \, 1, \, \cdots d-1\, ,  \nonumber \\
&&\eta=diag(-1,+1, \cdots , +1)\, .
\ee
 The covariant derivative is defined as a combination of the purely gravitational
covariant derivative and the gauge field $A_i=i B_i+A_i^a\lambda^a$, where $\lambda^a$ form the algebra of non-abelian transformations and $B_i$ is the abealian gauge field,
\be
\hat{\nabla}_i=\nabla_i +A_i\, .
\ee
The gravitational covariant derivative is defined as
\be
\nabla_k\psi=(\p_k+\frac{1}{2}\omega_k^{pq}\Sigma_{pq})\psi\, .
\ee
where $\Sigma^{pq}=\frac{1}{4}[\gamma^p,\gamma^q]$ and $\omega_k^{pq}$ is the spin connection\footnote{Also known as coefficients of Fock-Ivanenko \cite{Fock-Ivanenko}.} defined via the relation
\be
\nabla_i\,e_j^p=\p_i\,e_j^p-\Gamma^k_{ij}\,e_k^p+\omega_i\,^p\,_q\, e^q_i=0\, .
\ee
Defining $R_{ij}\,^{pq}=e^p_ke^q_\ell R_{ij}\,^{k\ell}$ one has\footnote{Note that our convention for the Riemann tensor differs by sign from the one used in \cite{Vassilevich:2003xt}.
On the other hand our convention for Ricci tensor and Ricci scalar agree with \cite{Vassilevich:2003xt}.} 
\be
R_{ij}\,^{pq}=\p_i\omega_j\,^{pq}-\p_j\omega_i^{pq}+[\omega_i,\omega_j]^{pq}\, .
\ee
A direct calculation shows that
\be\label{comm}
[\nabla_i,\nabla_j]=\frac{1}{4}R_{ij}\,^{pq}\gamma_p\gamma_q\, .
\ee
From this we may define the field strength tensor as
\be
\Omega_{ij}=[\hat{\nabla}_i,\hat{\nabla}_j] =F_{ij}+\frac{1}{4}R_{ij}\,^{pq}\gamma_p\gamma_q \, .
\ee
The Laplace type operator for Dirac theory is the square of Dirac operator,
\be
\Delta^{(\frac{1}{2})}\psi \equiv (i\gamma^k \hat{\nabla}_k)^2=-(\hat{\nabla}^2+E)\psi \, .
\ee
For $E$ one finds
\be
E=-\frac{1}{4}R +\frac{1}{4}[\gamma^i, \gamma^j] F_{ij}\, .
\ee

\subsection{Fermions in even and odd dimensions}
First we discuss the  Dirac gamma matrices in even and odd dimensions. 

\subsection{Even dimension $d$}
In even dimensions there is a unique representation for the Clifford algebra in terms of the $2^{\frac{d}{2}}\times 2^{\frac{d}{2}}$ unitary matrices. In even dimensions, the following unitary matrix anti-commutes with all gamma matrices and thus it can be used to introduce a chiral representation,
\be
\gamma^*=-i^{\frac{d-2}{2}}\gamma^0\gamma^1\cdots\gamma^{d-1}\, , \ \  (\gamma^*)^2=1\, .
\ee
Explicitly, for $d=2$
\be
\gamma^0=-i\sigma^1\, , \gamma^1=\sigma^2\, ,  \gamma^*=-\gamma^0\gamma^1=-i\sigma^1\sigma^2=\sigma^3\, ,
\ee
where $\sigma^i\, , \, i=1,\, 2,\, 3$ are the $2\times 2$ Pauli matrices.

For $d=4$ one has
\be
\gamma^0=
\begin{pmatrix}
0 & -i\mathbb{I}\\
-i\mathbb{I} & 0
\end{pmatrix}\, ,
\gamma^{1,2,3}=
\begin{pmatrix}
0 & -i\sigma^{1,2,3}\\
i\sigma^{1,2,3} & 0
\end{pmatrix}\, ,
\gamma^*=-i\gamma^0\gamma^1\gamma^2\gamma^3=
\begin{pmatrix}
\mathbb{I} & 0\\
0& -\mathbb{I}
\end{pmatrix}\, .
\ee
\subsubsection{Odd dimension $d$, doubling trick}
In odd dimensions for the standard representation $2^{\frac{d-1}{2}}\times 2^{\frac{d-1}{2}}$ for the gamma matrices the product
\be
\gamma^{d-1}_{\mp}=\mp i^{\frac{d-3}{2}}\gamma^0\gamma^1\cdots \gamma^{d-2}
\ee
belongs to the Clifford algebra by itself. Here, two different signs determine two unequal sets of Dirac matrices. 
Thus, in odd dimensions there does not exist a matrix, $\gamma^*$, sometimes called the chirality matrix,  that would anti-commute with all gamma matrices. For the reasons that will be clear shortly,
when we will discuss the appropriate mixed boundary conditions for the fermions, we would need such a matrix to exist.
This is the primary reason why we shall consider a doubled representation of gamma matrices obtained in a procedure sometimes referred  to as a doubling procedure.
The physical meaning of this doubling can be understood in this way that, for chirality to be meaningful, we need two distinguished spin states in odd dimensions, which is what the doubling procedure provides \cite{doubling}.
Following the trick we define
\be\label{dt}
\Gamma^{k}=\gamma^{k}\otimes\sigma^2 \ , \ k=0,1,\cdots\, , d-1\, ,
\lb{gammas}
\ee
where in the last step we choose $\gamma^{d-1}\equiv\gamma^{d-1}_-$. These new gamma matrices satisfy the Clifford algebra relations,
\be
\Gamma^k \Gamma^\ell +\Gamma^\ell \Gamma^k=2 \eta^{k\ell}\,.
\ee
The product of first $d-2$ gamma matrices now is not the same as $\Gamma^{d-1}$. The respective Dirac fermions have $2^{\frac{d+1}{2}}$ components, twice as the standard Dirac fermions.
 The other interesting feature is that now there exist two candidates for the chiral matrix,
 \be
  \Gamma^*_1= \mathbb{I} \otimes \sigma_1\, , \  \   \Gamma^*_2= \mathbb{I} \otimes \sigma_3\, , \  \   \ (\Gamma^*_1)^2=1\, , \ \  \ (\Gamma^*_2)^2=1\, .
 \ee
 Both these matrices are Hermitian and anticommute with all gamma matrices (\ref{gammas}). Thus, one may define two chiral type transformations,
 \be
 \psi\rightarrow e^{i\Gamma^*_1\alpha}\psi \ \ \ {\rm and } \  \  \  \psi\rightarrow e^{i\Gamma^*_2\beta}\psi \, .
 \lb{chiral}
 \ee
More generally, four matrices, $ \mathbb{I}$, $\Gamma^*_1$, $\Gamma^*_2$ and $[\Gamma^*_1, \Gamma^*_2]$, generate a unitary group of transformations.
More on this representation and its applications in physics can be found in \cite{d=3 fermions}.  

\vspace{1cm}
\noindent {\bf d=3}
\vspace{0.3cm}

In $3$ dimensions we start with
\be
\gamma^0=-i\sigma^1 \ \ , \ \ \gamma^1=\sigma^2\ , \ \ \gamma^2=-\gamma^0\gamma^1=\sigma^3\, .
\ee
Therefore
\be
\Gamma^0=
\begin{pmatrix}
 0 & -i\sigma^2 \\
 -i\sigma^2 & 0
 \end{pmatrix}\ \ , \ \ 
\Gamma^1=
\begin{pmatrix}
 0 & -i\sigma^2 \\
 i\sigma^2 & 0
 \end{pmatrix}
\ \ , \ \ 
\Gamma^2=
\begin{pmatrix}
 \sigma^2 & 0 \\
 0 & -\sigma^2
 \end{pmatrix}\ .
\ee
These matrices are Hermitian, except $\Gamma^0$ which is anti-Hermitian. As explained above  we define two different chirality matrices as
\be
\Gamma^*_1=\mathbb{I}_{2\times 2}\otimes\sigma^1=
\begin{pmatrix}
 \sigma^1 & 0 \\
 0 & \sigma^1
 \end{pmatrix}\, ,
\ee
and
\be
\Gamma^*_2=\mathbb{I}_{2\times 2}\otimes\sigma^3=
\begin{pmatrix}
 \sigma^3 & 0 \\
 0 & \sigma^3
 \end{pmatrix}\, .
\ee
These two Hermitian matrices anticommute with all of the gamma matrices we introduced above.
It can be easily checked that
\be
\Gamma^*_1\Gamma^*_2\Gamma^0\Gamma^1\Gamma^2=-i\mathbb{I}_{4\times 4}\, ,
\ee
and thus
\be
\tr(\Gamma^*_1\Gamma^*_2\Gamma^i\Gamma^j\Gamma^k)=-4i\epsilon^{ijk}\, .
\ee

\vspace{1cm}
\noindent {\bf d=5}
\vspace{0.3cm}

In five dimensions
\be
\gamma^0=
\begin{pmatrix}
0 & -i\mathbb{I} \\
-i\mathbb{I} & 0
\end{pmatrix} \ , \ 
\gamma^{1,2,3}=
\begin{pmatrix}
0 & -i\sigma^{1,2,3} \\
i\sigma^{1,2,3}& 0
\end{pmatrix}\, .
\ee
On the other hand
\be
\gamma^4=-i\gamma^0\gamma^1\gamma^2\gamma^3=
\begin{pmatrix}
\mathbb{I} & 0\\
0 & -\mathbb{I}
\end{pmatrix} \ .
\ee
Then using \eqref{dt} one can construct $\Gamma^a\, , \, a=0, \dots , \, 4$.
Two chirality matrices in five dimensions are 
\be
\Gamma^*_1=\mathbb{I}_{4\times 4}\otimes\sigma^1\, ,
\ee
and
\be
\Gamma^*_2=\mathbb{I}_{4\times 4}\otimes\sigma^3\, .
\ee
Then one finds that 
\be
\Gamma^*_1\Gamma^*_2\Gamma^0\Gamma^1\Gamma^2\Gamma^3\Gamma^4=\mathbb{I}_{8\times 8}\, ,
\ee
and thus the trace is
\be\label{t5}
\tr (\Gamma^*_1\Gamma^*_2\Gamma^i\Gamma^j\Gamma^k\Gamma^\ell\Gamma^m)=8\epsilon^{ijk\ell m}\, .
\ee

\subsection{Boundary conditions}

In the present paper we are interested in  situation when the spacetime $M$ has a boundary $\partial M$. 
We consider a spacelike boundary so that $n^2=n_k n^k=1$ for the normal vector $n=n^k \partial_k$.
Respectively, near the boundary one may separate the normal direction given by vector $n$ and the directions along the boundary
given by a basis of tangent vectors $t^k_a\partial_k\, , \ a=1, \dots , d-1$. The appropriate mixed boundary conditions were first formulated
by Gilkey and Branson  \cite{Branson-Gilkey}.
In our discussion of the boundary conditions to be imposed on the Dirac fermions we  follow the chapter 3 in the book 
of Fursaev and Vassilevich \cite{VF}  and  give some  necessary clarifications. 

For a differential operator of order $q$ one has to impose $q$ initial conditions, i.e. conditions at the initial constant time hyperurface.
If there is a second, final constant time hypersurface, the required boundary conditions are distributed between them so that at each component
of the boundary one has to impose $q/2$ conditions. The Dirac operator   $\hat{D}=i\gamma^k \hat{\nabla}_k $ is a first order operator so that in this case one has to impose the boundary conditions
on a half of fermionic components. Suppose that $\Pi_+$ is the projector which selects a half of the spinor components.  One can define
$\Pi_-=1-\Pi_+$ the projector on the other half. As projectors they satisfy the properties: $\Pi_\pm^2=\Pi_\pm$ and $\Pi_+\Pi_-=\Pi_-\Pi_+$.

A natural physical condition is to require that the normal component of the fermionic current  vanishes on the boundary (for simplicity we use here the Euclidean signature),
\be
\psi^\dagger \gamma^n \psi |_{\partial M}=0\, , \ \  \gamma^n=n_k \gamma^k\, .
\ee
This can be achieved by imposing the Dirichlet boundary condition
\be
\Pi_-\psi |_{\partial M}=0\, ,
\lb{Dirichlet}
\ee
provided the projector $\Pi_-$ satisfies certain commutation condition with $\gamma^n$.  To identify this condition, decomposing $\psi = \Pi_+\psi+\Pi_-\psi$
and using  that the projectors are unitary and assuming condition (\ref{Dirichlet}) we find
\be
\psi^\dagger \gamma^n \psi |_{\partial M}=\psi^\dagger \Pi_+ \gamma^n \Pi_+\psi |_{\partial M}=\eta\psi^\dagger \gamma^n \Pi_+\Pi_-\psi |_{\partial M}=0\, ,
\lb{conditions-b}
\ee
where we used the commutativity of the two projectors and assumed a relation $\Pi_+\gamma^n=\eta \gamma^n \Pi_-$. In order to determine the value of $\eta$ we apply once
again projector $\Pi_+$ to both sides of this relation and get equation $\eta^2=\eta$ that gives us $\eta=1$. 

The square of Dirac operator $\hat{D}^2$ is an operator of second order in derivative and logically one has to impose more conditions on the spinor components than just 
(\ref{Dirichlet}).  These conditions should be valid at least for the eigenvectors of the Dirac operator $\hat{D}\psi=\lambda \psi$.
Applying projector $\Pi_-$ to both sides of this equation and assuming the condition (\ref{Dirichlet}) we arrive at
\be
\Pi_- \hat{D}\psi |_{\partial M}=0\, .
\lb{Robin}
\ee
In order to simplify our discussion let us consider the flat spacetime and the boundary $\partial M$ to be a plane $x^n=\text{const.}$ while $\{ x^a \}$ are the coordinates on the boundary.
Then we may separate the normal and tangential components  in the Dirac operator, $\hat{D}=i\gamma^n \partial_n +i\gamma^a\partial_a$.
Then condition (\ref{Robin}) leads to
\be
\Pi_-(\gamma^n \partial_n+\gamma^a\partial_a)|_{\partial M}=\gamma^n\partial_n\Pi_+\psi|_{\partial M}+ \Pi_-\gamma^a\partial_a\psi |_{\partial M}=0\, .
\lb{commutator}
\ee
Since the last term contains derivatives along the boundary we expect this term to vanish due to (\ref{Dirichlet}). This is so provided a commutation relation
is valid, $\Pi_-\gamma^a=\gamma^a\Pi_-$ (again the consistency condition requires that a possible numerical pre-factor $\eta$ in this relation to be $1$).
Thus we arrive at the Robin type boundary condition on the second half of the spinor components,
\be
\partial_n\Pi_+\psi|_{\partial M}=0\, .
\lb{Robin2}
\ee
We have also found the commutation relations between the projectors and the gamma matrices,
\be
\Pi_+\gamma^n =\gamma^n \Pi_- \, , \   \   \  \Pi_-\gamma^a=\gamma^a \Pi_-\, .
\ee
Representing $\Pi_\pm =\frac{1}{2}(1\pm \chi)$ we find that $\chi$ has to anti-commute with $\gamma^n$ and commute with $\gamma^a$.
This matrix $\gamma$ can be constructed as follows
\be
\chi=i\gamma^* \gamma^n\, ,
\ee
where $\gamma^*$ anti-commutes with all gamma matrices. This is so called the chirality matrix.

In  the case of a curved spacetime with a boundary $\partial M$ characterized by non-trivial extrinsic curvature $K$, the commutations that we performed in
(\ref{commutator}) are more involved and they are performed in Appendix  \eqref{appC}. The respective Robin type boundary condition that generalises  (\ref{Robin2}) is
\be
(\hat{\nabla}_n-S)\Pi_+\psi=0\, , \  \  \  S=-\frac{1}{2}K\Pi_+\, ,
\lb{mixed}
\ee
where $K$ is the trace of the extrinsic curvature which we briefly introduce in subsection \eqref{EG}.
We note that the mixed boundary conditions (\ref{mixed}) are conformal invariant.

\vspace{0.5 cm}
\noindent{\bf Even dimensions.} In even dimensions, the mixed boundary conditions on the boundary $\p M$ will be realized as
\be\label{mbc}
\Pi_-\psi\vert_{\p M}\oplus(\hat{\nabla}_n-S)\Pi_+\psi\vert_{\p M}=0\, .
\lb{bceven}
\ee
We remind again that for Dirac spinors $\Pi_{\pm}=\frac{1}{2}(1\pm i\gamma^*\gamma^n)$, where $\gamma^*$ is the chirality matrix and $\gamma^n=n_k\gamma^k$. 
\vspace{0.5cm}

\noindent{\bf Odd dimensions.} In the standard representation for the gamma matrices of dimension $2^{\frac{d-1}{2}}\times 2^{\frac{d-1}{2}}$ there exists no
a matrix that would anti-commute  with all gamma matrices. This poses a problem for formulation of appropriate boundary conditions for the Dirac spinors.
This forces us to use the other known representation for the gamma matrices that we discussed in section 2.3.1.
Boundary conditions in odd dimensions are obtained in the same way as in even dimensions and by replacing $\gamma^k$ with $\Gamma^k$ and $\gamma^*$ with one of the two chirality matrices, for instance $\Gamma^*_1$. So in odd dimensions we set
\be
\chi=\Pi_+-\Pi_-= i\Gamma^*_1\Gamma^n\, .
\ee
One should note that the duplication due to the existence of two nonequivalent representations in odd dimensions is very crucial for properly setting the boundary conditions on fermions. Otherwise, the set of boundary conditions would be ill-defined and over-restricted in such cases. As a result of this duplication, all traces will be doubled when we take the trace of the various terms contributing to the boundary conformal anomalies in odd dimensions. Applying these boundary conditions, we calculate the boundary anomaly for two explicit examples of $d=3$ and $d=5$, in the following.
In odd dimension $d$, we thus have
\be
\tr \mathbb{I}=2d_s\ \ , \ \ \tr\Pi_{\pm}=d_s \, ,
\ee
where $d_s=2^{\frac{d-1}{2}}$. 

Notice that in any (odd or even) dimension
one has that
\be
\tr \gamma=0\, .
\ee

\vspace{0.5 cm}
\noindent{\bf Chiral transformations.} For a massless Dirac field one may introduce a chiral transformation $\psi\rightarrow e^{i\alpha\gamma^*}\psi$ where $\gamma^*$ is a chirality matrix.
The Dirac action is invariant under such transformation. In even dimension $d$, the boundary conditions (\ref{bceven}) are also invariant. 
In the case of odd dimension $d$ there exist two chirality matrices $\Gamma^*_1$ and $\Gamma^*_2$ and respectively two
possible choices for the chiral transformations (\ref{chiral}). If we choose matrix $\Gamma^*_1$ to define the projectors $\Pi_+$ and $\Pi_-$ in the mixed boundary condition then
these conditions are invariant under the chiral transformations generated by the other chirality matrix\footnote{The other possibility is to choose $\Gamma^*_1$ as the chirality matrix. Then this is similar to what one has in the even dimensions, for instance $d=4$. In particular, one has same difficulty (and same resolution)  in defying the invariant boundary conditions that we discuss at the end of section 3. As far as we can see, the respective chiral anomaly vanishes both in $d=3$ and $d=5$ dimensions. That is why we do not consider this case here.}, $\Gamma^*_2$. The invariance under chiral transformations 
means that the current $j^i_{A}=\bar{\psi}\gamma^i \gamma^*\psi$ is conserved, $\nabla_i j^i_{A}=0$. The conservation is violated in quantum theory that leads to
the  quantum chiral anomaly. Below we shall compute the anomaly in dimensions $d=3, \, 4 $ and $5$.

\subsection{Extrinsic geometry}\label{EG}
To complete the discussion in this section, let us briefly review the external geometry. External geometry is about how a boundary is embedded  in a manifold. The characteristic measure of this geometry is the extrinsic curvature tensor, or as it is often called in mathematical texts, the second fundamental form. This tensor will be defined as
\be
K_{ij}=h^k_ih^\ell_j  \nabla_{(k}n_{l)}\, ,
\ee
where $n^i$ is the unit normal vector on the boundary and $h^i_j=\delta^i_j-n^in_j$ defines the projection on the boundary. This tensor is symmetric by construction, with no component in the normal direction, $n^iK_{ij}=0$. 
If it is preferred to consider the extrinsic curvature as a tensor living on the boundary, $\p M$ the following pullback can be calculated
\be
K_{ab}=t^i_at^j_b\, K_{ij}\, .
\ee
Accordingly, one can define the induced metric on, or the first fundamental form of the boundary
\be
h_{ab}=t^i_at^j_b\,  h_{ij}=t^i_at^j_b\,  g_{ij}\, .
\ee
We denote the trace of this tensor, which frequently appears in our equations, by $K=K_i^i=K^a_a$. 

The covariant derivative defined with respect to the intrinsic metric $h_{ab}$ is denoted by $\bar{\nabla}_a$ and the respective curvature by $\bar{R}$, $\bar{R}_{ab}$ and $\bar{R}_{abcd}$.
The relations between the intrinsic curvature of the boundary and the curvature in the $5$ dimensional space-time are given by the Gauss-Codazzi identities presented in  Appendix \eqref{Gauss-Codazzi}. 

Under the infinitesimal  conformal transformations $\delta g_{ij}=2\sigma g_{ij}$, $\delta n_{i}=\sigma n_{i}$ the Weyl tensor transforms as $\delta W_{ijkl}=2\sigma W_{ijkl}$.
The extrinsic curvature transforms as follows
\be
\delta{K}_{ab}=\sigma K_{ab}+\gamma_{ab}\nabla_n\sigma \, , \ \ \ \delta K=-\sigma K+4\nabla_n\sigma\, , \ \ \ \delta\hat{K}_{ab}=\sigma\hat{K}_{ab}\, ,
\lb{2}
\ee
where $\nabla_n=n^i\nabla_i$ and $\hat{K}_{ab}=K_{ab}-\frac{1}{d-1}h_{ab}K$ is  the trace-free part of the extrinsic curvature tensor.
The basic conformal tensors are, thus, the bulk Weyl tensor $W_{ijkl}$ and the trace free extrinsic curvature of the boundary $\hat{K}_{ab}$. The intrinsic Weyl tensor of the boundary metric is expressed in terms of
the bulk  Weyl tensor and the extrinsic curvature by means of the Gauss-Codazzi relations.

\section{Conformal and chiral anomalies}
The quantization of the fermionic field in a fixed gravitational and gauge field background leads to the quantum effective action,
\be
W_Q=-\frac{1}{2}\ln \det \hat{D}^2\, ,
\ee
expressed in terms of determinant of the square Dirac operator $\hat{D}^2$. It can be calculated by using the heat kernel $K(x,x')=\langle x| e^{-s\hat{D}^2} |x'\rangle$,
\be
W_Q=\frac{1}{2}\int_{\epsilon^2}^\infty \frac{ds}{s}\Tr K(\hat{D}^2,x,x, s)\, ,
\ee
where the trace includes also integration over $x$. The trace of the heat kernel is characterized by its small $s$ expansion,
\be
\Tr K(\hat{D}^2, s)=\sum_{p=0}a_p(\hat{D}^2)s^{\frac{(p-d)}{2}}\, , \  \  \  s\rightarrow 0
\ee
where $a_p(\hat{D}^2)$ are the heat kernel coefficients that are represented by the bulk and boundary integrals,
\be
a_p(\hat{D}^2)=\int_M \tr A_p (x)+ \int_{\partial M} \tr B_p(x)\, ,
\ee
where the trace is taken over spinor and group indexes, $A_p(x)$ and $B_p(x)$ are local invarinats 
constructed from the curvature, gauge fields 
and the extrinsic curvature, the latter invariants appear in the boundary term $B_p(x)$. 

The invariance of the classical theory under  local conformal transformations means that the respective classical stress energy tensor is traceless.
This property however is violated in the quantum theory that manifests in the conformal anomaly.
For a quantum Dirac field the integrated conformal anomaly  in dimension $d$ is determined by the coefficient $a_d$, see \cite{RS},  \cite{Vassilevich:2003xt},
\be
\int_{{\cal M}_d}\la T_{ij}\ra g^{ij}=-a_d(\hat{D}^2)\, .
\lb{anomaly}
\ee
The minus sign here is due to anti-commutativity of the fermion fields in the path integral.

Provided there exists a chirality matrix $\gamma^*$, the theory classically possesses a conserved axial current $j^i_{A}=\bar{\psi}\gamma^i\gamma^*\psi$, 
$\nabla_i j^i_A=0$.
In quantum theory the conservation is modified by a quantum anomaly.  The local form of the anomaly in terms of the heat kernel coefficients was found in earlier works, see for instance 
 \cite{RS},  \cite{Vassilevich:2003xt}). When there are both bulk and boundary contributions to the anomaly it can be presented as follows, 
\be
\nabla_i \la j^i_{A} (x)\ra=-2i \left( \tr (\gamma^* A_d(x))+\tr (\gamma^* B_d(x))\delta_{\partial M}  \right)\, .
\lb{chiralanomaly}
\ee

In dimension $d=4$ one has a difficulty since the boundary conditions (\ref{bceven}) are not invariant under the chiral transformations since the chiral matrix $\gamma^*$ anti-commutes rather than commutes with $\chi=\Pi_+-\Pi_-$.  Resolution of this difficulty leads to a generalisation of the boundary conditions and the matrix $\chi$. This issue was earlier discussed in  \cite{Gilkey:2005qm}, \cite{Beneventano:2003hv}, \cite{Ivanov:2021yms}. One introduces a family of the boundary conditions (known as the chiral bag boundary conditions) with $\chi (\theta)$ defined as
\be
\chi (\theta)=i\gamma^* \gamma^n e^{i\gamma^* \theta}\, ,
\ee 
where $\theta$ is a parameter. Under the chiral transformations $\psi\rightarrow e^{i\gamma^*\phi}\psi$ it transforms as
\be
\chi (\theta)\rightarrow e^{-i\gamma^*\phi}\chi (\theta) e^{i\gamma^*\phi}=\chi(\theta+2\phi)\, .
\ee
The boundary conditions remain invariant if one transforms $\theta\rightarrow \theta-2\phi$. This transformation is even more natural if the parameter of transformation is a local function
of coordinates. Then $\theta$ is a function of the coordinates and to have the theory invariant one introduces axial gauge field that compensates the gradients of $\phi(x)$.
This direction was followed in \cite{Ivanov:2021yms}. We do not consider here this generalisation and keep parameter $\theta$ to be constant. Then one has to compute the heat kernel of the
Dirac operator subject to the boundary conditions parametrised by $\theta$. The heat kernel coefficients then become the rather non-trivial functions of $\theta$.
A remarkable fact, however, proved in \cite{Gilkey:2005qm}, \cite{Beneventano:2003hv}, \cite{Ivanov:2021yms} is  that the heat kernel coefficient $a_4 (\hat{D}^2,\chi(\theta))$
is independent of  parameter $\theta$. For the chiral anomaly this means that
\be
\partial_\theta\Tr (\gamma^* a_4(\hat{D}^2, \chi(\theta)))=0\, .
\ee
This property justifies that in what follows we do the computation of the anomaly  for $\theta=0$.

The other technical but important remark is the following.  The computation of the chiral anomaly with the help of (\ref{chiralanomaly}) requires the asymptotic expansion
$\tr F(x) e^{-s\hat{D}^2}$ with $F(x)$ being a  matrix while the results available in \cite{Vassilevich:2003xt} and \cite{Branson:1999jz} are given for the case when $F(x)$ is unity matrix multiplied by a function.  So that these results can not be directly applicable\footnote{We thank the anonymous referee for rasing this issue.}. The difference is essential, there may appear terms that do not commute with $F(x)$. Some such terms due to different ordering of the operators in the heat kernel expansion have been computed in \cite{Marachevsky:2003zb}.
It should be noted that this problem does not arise in odd dimensions. Indeed, in this case $F(x)=\Gamma^*_2$ commutes with all operators that may enter the heat kernel expansion, $E$, $\Omega_{ij}$, $\chi$ and $\nb_a\chi$. So that the order of operators is not important in this case and one may use the asymptotic expansion obtained in  \cite{Vassilevich:2003xt} and \cite{Branson:1999jz}.  In dimension $d=4$ the situation is rather different. The chirality matrix $\gamma^*=\gamma_5$ commutes with $E$ and $\Omega_{ij}$ but anti-commutes with
$\chi$ and $\nb_a\chi$. This means that in the bulk one may still use the expansion found in  \cite{Vassilevich:2003xt} and \cite{Branson:1999jz}. However, on the boundary 
the order of operators becomes important. Luckily, the boundary terms in the asymptotic expansion  that contribute to the chiral anomaly in $d=4$ were previously computed in \cite{Marachevsky:2003zb}. We return to this issue in section 5.2 when compute the chiral anomaly in $d=4$.

\section{Boundary  anomaly for Dirac spinors in $d=3$ dimensions }

\subsection{Conformal anomaly}
In $d=3$ dimensions there are two boundary conformal invariants. One invariant is the integrated Euler density
\be
E_2=\int_{\p M_3}\bar{R}=\int_{\p M_3}(R-2R_{nn}-\Tr K^2+K^2)\, ,
\lb{E2}
\ee
where in the second equality we used the Gauss-Coddazzi relations, and the other is
\be
I_1=\int_{\p M_3}\Tr\Kh^2=\int_{\p M_3}(\Tr K^2-\frac{1}{2}K^2)\, ,
\lb{I1}
\ee
where $\Kh_{ab}$ is the traceless part of the extrinsic curvature tensor. 
The general form of the conformal anomaly in $d=3$ dimensions then is 
\be
\int_{{\cal M}_d}\la T_{ij}\ra g^{ij}=-\frac{a}{384\pi} E_2+\frac{c}{256\pi} I_1\, .
\lb{d=3 anomaly}
\ee
For a conformal scalar field  the conformal charges were computed in \cite{Solodukhin:2015eca}. In the present normalisation
 $(a=-1,c=+1)$ for a scalar field satisfying the Robin boundary conditions.

Computationally, we use equation (\ref{anomaly}) that expresses anomaly in terms of the heat kernel coefficient,
\be
\int_{{\cal M}_d}\la T_{ij}\ra g^{ij}=-a_3 
\lb{anomaly-heat}
\ee
of the Dirac operator. The general form of this coefficient is given in (\ref{a3}). Notice that in $d=3$ one defines $\chi=i\Gamma^*_1\Gamma^n$.
 The various constituent terms of $a_3$ and their contributions  are listed below
\begin{center}
\begin{tabular}{|c c c|}
 \hline
 various terms & extended forms & contribution to anomaly \\ 
  \hline
 $R$ & $R$ & $\tr(16\chi)=0$ \\ 
 $E$ & $-\frac{1}{4}R+\frac{1}{2}F_{ij}\Gamma^i\Gamma^j$ & $\tr(96\chi)=0,  \tr(\chi\Gamma^i\Gamma^j)=0$ \\ 
$R_{nn}$ & $R_{nn}$ & $\tr(-8\chi)=0$ \\ 
$\Tr K^2$ & $\Tr K^2$ & $\tr(2\Pi_++10\Pi_-)=24$ \\ 
$K^2$ & $K^2$ & $\tr(13\Pi_+-7\Pi_-)=12$ \\ 
$SK$ & $-\frac{1}{2}K^2$ & $\tr(96\Pi_+)=192$ \\ 
$S^2$ & $\frac{1}{4}K^2$ & $\tr(192\Pi_+^2)=384$ \\
$\tr (\nabla_a\chi\nabla^a\chi)$ & $4\Tr K^2$ & $-12$\\
 \hline
\end{tabular}
\end{center}
Note that in the third column we have dropped the overall factor $\frac{1}{384(4\pi)}$.

Collecting all terms one finds,
\be
a_3=\frac{1}{384(4\pi)}\int_{\p M_3}(-24\Tr K^2+12K^2)\, ,
\lb{result d=3}
\ee
and this should be re-written as a combination of the conformal invariants (\ref{E2}) and (\ref{I1}). Matching the coefficients between  (\ref{d=3 anomaly}) and 
(\ref{anomaly-heat}) we get the following algebraic equations
\begin{equation*}
\begin{split}
&R\ : \ a=0\, ,\\
&R_{nn}\ : \ a=0\, ,\\
&\Tr K^2\ : \ a+\frac{3}{2}c_1=6\, ,\\
&K^2\ : \ -a-\frac{3}{4}c_1=-3\, .\\
\end{split}
\end{equation*}
Solving these equations we find
\be
a=0\, ,  \  \  \   c=4\, .
\ee
Value of the charge $c=4$ corresponds to number of components of the spinor in the doubling representation. Notice that in terms of the charges
the $d=3$ spinor in this representation can be viewed as equal number of conformal scalars satisfying the Dirichlet $(a=1, c=+1)$ and  
conformal Robin $(a=-1,c=+1)$ boundary conditions. Note that in \cite{Fursaev:2016inw} the conformal anomaly due to the Dirac field in the standard $2\times 2$
representation was earlier computed.  Taking however our notes on the difficulties with imposing consistently the boundary conditions in the standard
representation that calculation was not fully eligible. Notice also that the gauge field $A_i$ does not make a contribution to the conformal anomaly in $d=3$ dimensions.

\subsection{Chiral anomaly in $d=3$ dimensions}
The chiral transformations that preserve the boundary conditions are defined with respect the chirality matrix $\Gamma^*_2$. Computing the chiral anomaly
according to (\ref{chiralanomaly}) we have to compute the trace in the heat kernel coefficient with matrix $\Gamma^*_2$. Most of the traces vanish, for instance
$\tr(\Gamma^*_2\chi)=0$. The only non-vanishing term is $\tr(\Gamma^*_2\chi\Gamma^i\Gamma^j)=-4\epsilon^{nij}$. 
The chiral anomaly then is due to the gauge fields,
\be
\tr(\Gamma^*_2, B_3(x))=-\frac{1}{8\pi} \epsilon^{nij}F_{ij}\, .
\ee
One can identify $\epsilon^{nij}=\epsilon^{ij}$, the intrinsic epsilon tensor defined on the boundary $\partial M_3$.
In this expression we are supposed also take trace over the group indexes. Taking that the non-abelian generators are traceless only the abelian component survives. 
The chiral anomaly then in
$d=3$ is entirely due to abelian gauge field $B^i$. The local chiral  anomaly is thus due to a boundary term,
\be
\nabla_i \la j^i_{A} \ra =-\frac{1}{2\pi} \epsilon^{ij}\partial_i B_j \, \delta_{\partial M_3}\, .
\ee
Notice that the anomaly has only a boundary contribution that is precisely of the form of a chiral anomaly in two dimensions .

\section{Anomaly in $d=4$ dimensions}
\subsection{Conformal anomaly  in $d=4$ dimensions }
In $d=4$ dimensions the integrated conformal anomaly has both bulk and boundary parts as was noticed in \cite{Fursaev:2015wpa}.
The computation we are about to perform in this section is the one already done by Fursaev \cite{Fursaev:2015wpa}. We of course reproduce his result
and include it here only for the reasons of completeness of our consideration.
In total, there are four possible conformal invariants so that the integrated anomaly reads
\be
\int_{{\cal M}_d}\la T_{ij}\ra g^{ij} =-\frac{a}{180}\chi[M_4]+\frac{b}{1920\pi^2}\int_{M_4}W_{ijkl}^2+\frac{c_1}{240\pi^2}\int_{\p M_4} \Kh^{ab}W_{anbn}+\frac{c_2}{280\pi^2}\int_{\p M_4}\Tr\Kh^3
\lb{4-anomaly}
\ee
where we excluded the contribution due to the gauge fields. These contributions will be discussed later separately.
The values for the charges $a$ and $b$ for free fields are well known. For a Dirac field one has $a=11$ and $b=6$.
We here focus on the boundary terms.

The general expression for heat kernel coefficient $a_4$, is given by (\ref{a4}).
 The various constituent boundary terms of $a_4$ and their contributions  are listed below. 
\begin{center}
\begin{tabular}{|c c c|}
 \hline
 various terms & extended forms & contribution to anomaly \\ 
  \hline
 $\nabla_n R$ & $\nabla_nR$ & $\tr(12\mathbb{I}-42\Pi_++18\Pi_-)=0$ \\ 
 $\nabla_n E$ & $-\frac{1}{4}\nabla_nR$ & $\tr(60\mathbb{I}-240\Pi_++120\Pi_-)=0$ \\ 
$K^{ab}R_{ab}$ & $K^{ab}R_{ab}$ & $\tr(-4\mathbb{I})=-16$ \\ 
$K^{ab}R_{anbn}$ & $K^{ab}R_{anbn}$ & $\tr(16\mathbb{I})=64$ \\ 
$KR_{nn}$ & $KR_{nn}$ & $\tr(-4\mathbb{I})=-16$ \\ 
$KR$ & $KR$ & $\tr(20\mathbb{I})=80$ \\ 
$KE$ & $-\frac{1}{4}KR$ & $\tr(120\mathbb{I})=480$ \\ 
$SR$ & $-\frac{1}{2}KR$ & $\tr(120\Pi_+)=240$ \\ 
$SE$ & $\frac{1}{8}KR$ & $\tr(720\Pi_+)=1440$ \\ 
$\Tr K^3$ & $\Tr K^3$ & $\tr(\frac{224\Pi_++320\Pi_-}{21})=\frac{1088}{21}$ \\ 
$K\Tr K^2$ & $K\Tr K^2$ & $\tr(\frac{168\Pi_+-264\Pi_-}{21})=-\frac{64}{7}$ \\ 
$S\Tr K^2$ & $-\frac{1}{2}K\Tr K^2$ & $\tr(48\Pi_+)=96$ \\ 
$K^3$ & $K^3$ & $\tr(\frac{280\Pi_++40\Pi_-}{21})=\frac{640}{21}$ \\ 
$S K^2$ & $-\frac{1}{2}K^3$ & $\tr(144\Pi_+)=288$ \\ 
$S^2 K$ & $\frac{1}{4}K^3$ & $\tr(480\Pi^2_+)=960$ \\ 
$S^3$ & $-\frac{1}{8}K^3$ & $\tr(480\Pi_+^3)=960$ \\
$\tr (\chi\nabla^a\chi\Omega_{an})$ & $2K^{ab}R_{anbn}$ & $-60$\\
$K\tr (\nabla_a\chi\nabla^a\chi)$ & $4K\Tr K^2$ & $-12$\\
$K^{ab}\tr (\nabla_a\chi\nabla_b\chi)$ & $4\Tr K^3$ & $-24$\\
$\tr (\nabla_a\chi\nabla^a\chi S)$ & $-K\Tr K^2$ & $-120$\\
 \hline
\end{tabular}
\end{center}
Above  in the contribution to anomaly we drop the overall factor $\frac{1}{360(4\pi)^2}$.
So for  the anomaly  
\be
\int_{{\cal M}_4}\la T_{ij}\ra g^{ij} =-a_4\, ,
\ee
and focusing only on the boundary terms we find
\be
\begin{split}
a_4&=\frac{1}{360(4\pi)^2}\int_{\p M_4}\Big(-56K^{ab}R_{anbn}-16K^{ab}R_{ab}-16KR_{nn}+20KR\\
&-\frac{928}{21}\Tr K^3+\frac{104}{7}K\Tr K^2+\frac{136}{21}K^3\Big)\, .
\end{split}
\ee
Matching the coefficients with the general expression (\ref{4-anomaly}) one arrives at the following algebraic equations
\begin{equation*}
\begin{split}
&K^{ab}R_{anbn}\ : \ -8a-24c_1 =56\, ,\\
&K^{ab}R_{ab}\ : \ 8a-12c_1=16\, ,\\
&KR_{nn}\ : \ 8a-12c_1=16\, ,\\
&KR\ : \ -4a+4c_1=-20\, ,\\
&\Tr K^3\ : \ -\frac{16}{3}a+\frac{144}{7}c_2 =\frac{928}{21}\, ,\\
&K\Tr K^2\ : \ 8a-\frac{144}{7}c_2=-\frac{104}{7}\, ,\\
&K^3\ : \ -\frac{8}{3}a+\frac{32}{7}c_2=-\frac{136}{21}\, ,\\
\end{split}
\end{equation*}
solving which we find
\begin{table}[H]
\begin{center}
\renewcommand{\baselinestretch}{2}
\medskip
\bigskip
 \begin{tabular}[H]{ | c | c | }
    \hline
  \text{Conformal charges} & \text{Mixed b.c.}\\
  \hline
  $a$ & $11$ \\
  \hline
    $b$ & $6$ \\
  \hline
$c_1$ & $6$\\
  \hline
$c_2$ & $5$  \\
  \hline
\end{tabular}
\renewcommand{\baselinestretch}{1}
\end{center}
\end{table}

\noindent  The value of $b$ comes from
matching the bulk terms and is well known. These values agree with those found in \cite{Fursaev:2015wpa}.

\vspace{0.5 cm}

Let us now discuss the contribution to the bulk and boundary conformal anomaly due to the gauge fields.
In the bulk the possible non-trivial contribution comes from the terms $\tr \Omega_{ij}\Omega^{ij}=4F_{ij} F^{ij}$ and  $\tr E^2=
\frac{1}{4}F_{ij}F_{kl}\tr \gamma^i \gamma^j\gamma^k\gamma^l=-2 F_{ij}F^{ij}$. Notice that in these expressions one also supposes to take trace over the group indexes, we omit it here to avoid additional confusion with traces.
Among the boundary terms there are several terms 
that depend either on $E=\frac{1}{2}F_{ij}\gamma^i\gamma^j$  or $\Omega_{ij}=F_{ij}$ (we here focus only on dependence on the gauge fields) 
that potentially may produce a contribution to anomaly due to the gauge fields. However, all these terms identically vanish.
For instance, $\tr (\Pi_+E)=0$ and $\tr(\chi\bar{\nabla}_a\chi \Omega^a_{\  n})=0$.
So we conclude that the only contribution to the conformal anomaly due to the gauge fields is in the bulk,
\be
\int_{{\cal M}_4}\la T_{ij}\ra g^{ij} =\frac{1}{24\pi^2}\int_{M_4} F_{ij}F^{ij}
\, ,
\ee
where  the trace over the group indexes of the gauge fields is assumed to be taken.
This anomaly is of course well known.

\subsection{Chiral anomaly in $d=4$ dimensions}
First we discuss purely gravitational part in the chiral anomaly.
We choose present result in the integrated form,
\be
\int_{M_4}\nabla_i \la j^i_{A} \ra=-2i (\int_{M^4}\tr (\gamma^* A_4(x))+\int_{\p M_4}\tr (\gamma^*B_4(p)))\, .
\lb{lm}
\ee
Analysis shows that only two terms are non-vanishing. One term is in the bulk, $\tr (\gamma^*\Omega_{ij}\Omega^{ij})$ and the other on the boundary
$\tr(\gamma^*\nb_a\chi\Omega^a_{\ n})$. Taking that $\Omega_{ij}=\frac{1}{4}R_{ijkl}\gamma^k\gamma^l$ and that $\nb_a\chi=i\gamma^*\gamma^b K_{ab}$
(see appendix) we find using that $\tr(\gamma^* \gamma^i\gamma^j\gamma^k\gamma^\ell)=-4i\epsilon^{ijk\ell}$,
\be
&&\tr (\gamma^*\Omega_{ij}\Omega^{ij})=-\frac{i}{4}\epsilon^{klmn}R_{ijkl}R^{ijmn}\, , \nonumber \\
&&\tr(\gamma^*\chi\nb_a\chi\Omega^a_{\ n})=i\epsilon^{nbcd}R_{nacd} K_{ab}\, .
\ee
The first term contributes to the anomaly in the bulk and the second term to the anomaly on the boundary. As we discussed earlier in the paper, one can not directly use 
the boundary coefficient $B_4$ since it was obtained in the assumption that it is contracted with a unity operator $F(x)$ that commutes with all other operators 
that appear in the asymptotic expansion. In (\ref{lm}) matrix $F(x)=\gamma^*$ commutes with $E$ and $\Omega_{ij}$ but anti-commutes with $\chi$ and $\nb_a\chi$
that appear on the boundary. Therefore, one can not directly use $B_4$ given in appendix since there may appear several terms with different ordering of operators
that replace a single term in the commutative case. It is interesting that precisely the terms that we need for computing the chiral anomaly were found in \cite{Marachevsky:2003zb},
\be
\tr \left[F(x)\left(-18 \chi \nb^a\chi\Omega_{an}-12\nb^a \chi\Omega_{an}\chi-18\Omega_{an}\chi\nb^a\chi+12\chi\Omega_{an}\nb^a\chi \right)\right]\, .
\ee
In commutative case ($F(x)$ is proportional to unity matrix) these terms combine to $-60\tr (F(x)\chi \nb^a\chi\Omega_{an})$ that comes from $B_4$ in appendix.
For $F(x)=\gamma^*$ these terms  combine to produce 
\be
-12\tr (\gamma^*\chi\nb^a\chi\Omega_{an})\, ,
\ee
where we used that $\chi$ and $\nb_a\chi$ anti-commute.

Combining these elements of the calculation we conclude that
\be
\int_{M_4}\nabla_i \la j^i_{A} \ra=-\frac{1}{384\pi^2}\left[\int_{M_4}\epsilon^{k\ell pq}R_{ijk\ell}R^{ij}\,_{pq}+\frac{8}{5}\int_{\p M_4}\epsilon^{nabc}K_a^dR_{ndbc}\right]\, .
\lb{chiral-4-1}
\ee
Using \eqref{GC2} we can rewrite the second integrand as follows
\be
\epsilon^{nabc}K_a^dR_{ndbc}=-2\epsilon^{nabc}K^d_a\nb_cK_{bd}\, ,
\ee
therefore
\be
\int_{M_4}\nabla_i \la j^i_{A} \ra=-\frac{1}{384\pi^2}\left[\int_{M_4}\epsilon^{k\ell pq}R_{ijk\ell}R^{ij}\,_{pq}-\frac{16}{5}\int_{\p M_4}\epsilon^{abc}K^d_a\nb_cK_{bd}\right]\, ,
\lb{chiral-4}
\ee
where in the boundary term we use intrinsic $\epsilon^{abc}=\epsilon^{nabc}$.
The first term here is $\frac{1}{12}P$, where $P$ is the Pontryagin number defined as
\be
P=\frac{1}{32\pi^2}\int_{M_4}\epsilon^{k\ell pq}R_{ijk\ell}R^{ij}\,_{pq}\, .
\ee
The boundary term in the chiral anomaly (\ref{chiral-4-1}), (\ref{chiral-4}) is new. At the moment it is not clear whether it should always be combined with the Pontryagin term or it is an independent
term. 
 
 The other remarks is that the chiral anomaly has a structure similar to that of the parity anomaly, see \cite{Kurkov:2018pjw}, although the relative coefficients are different.
The quantum origin of  these two anomalies is however quite different, see  discussion in \cite{Ivanov:2021yms}. For other recent works on parity anomaly see 
 \cite{Witten:2019bou} - \cite{Fresneda:2023wub}.

It is interesting to note that the boundary term (\ref{chiral-4-1}) is conformal invariant. Indeed it can be expressed in terms of the trace-free extrinsic curvature $\hat{K}_{ab}$ and the Weyl tensor,  $\epsilon^{nabc}K_a^dR_{ndbc}=\epsilon^{nabc}\hat{K}_a^dW_{ndbc}$. 

Our last comment in this section concerns the possible contribution of the gauge fields to the chiral anomaly.
Analysis shows that there is only one term in the heat kernel that contributes to the anomaly. This is a bulk term 
\be
\tr (\gamma^* E^2)=\frac{1}{4}F_{ij}F_{k\ell}\tr (\gamma^*\gamma^i\gamma^j\gamma^k\gamma^\ell)=-i\epsilon^{ijk\ell}F_{ij}F_{k\ell}\, .
\ee
We conclude that a contribution due to the gauge fields to the chiral anomaly is given by
\be
\int_{M_4}\nabla_i\langle j^i_A \rangle =-\frac{1}{16\pi^2}\int_{M_4}\epsilon^{ijk\ell}F_{ij}F_{k\ell}\, .
\ee
This is a known result in the literature. No boundary term due to the gauge fields in the anomaly  has been found.

\section{Anomaly  in $d=5$ dimensions }

\subsection{Boundary conformal anomaly in $d=5$ dimensions}
The conformal anomaly in five dimensions is entirely due to the boundary terms,
\be
\int_{{\cal M}_4}\la T_{ij}\ra g^{ij} =\frac{1}{5760(4\pi)^2}\int_{\p M_5}(a E_4+\sum_{k=1}^8 c_k I_k)\, .
\lb{anomaly5}
\ee
$E_4$ is the Euler density integrated over boundary $\p M_5$, $\chi[\p M_5]=\frac{1}{32\pi^2}E_4$ is the Euler number of the boundary.
$\{ I_k\, , \, k=1, \dots , 8\}$ is a set of eight conformal invariants. Altogether they form a complete basis of boundary conformal invariants 
in five dimensions. For convenience below we give exact expressions for all these invariants.

\subsubsection{Boundary invariants in $d=5$ dimensions}
We start our discussion of conformal anomaly in $d=5$ dimensions by recalling the complete list of boundary conformal invariants in five dimensions
\cite{FarajiAstaneh:2021foi},
\be
\begin{split}
&E_4=\int_{\p M_5}(\bar{R}_{acbd}^2-4\bar{R}_{ab}^2+\bar{R}^2)\\
&=\int_{\p M_5}\Big(R^2_{acbd}-4R^2_{ab}+R^2-4R_{anbn}^2+8R^{ab}R_{anbn}+4R^2_{nn}-4RR_{nn}+4K^{ab}K^{cd}R_{acbd}\\&+8(K^2)^{ab}R_{ab}-8KK^{ab}R_{ab}-2\Tr K^2R+2K^2R
-8(K^2)^{ab} R_{anbn}+8KK^{ab}R_{anbn}\\
&+4\Tr K^2R_{nn}
-4K^2R_{nn}-6\Tr K^4+8K\Tr K^3+3(\Tr K^2)^2-6K^2\Tr K^2+K^4\Big)\, ,
\end{split}
\lb{3}
\ee
\be
I_1=\int_{\p M_5}(\Tr\Kh^2)^2=\int_{\p M_5}[(\Tr K^2)^2-\frac{1}{2}K^2\Tr K^2+\frac{1}{16}K^4]\, ,
\ee
\be
I_2=\int_{\p M_5}\Tr \Kh^4=\int_{\p M_5}(\Tr K^4-K\Tr K^3+\frac{3}{8}K^2\Tr K^2-\frac{3}{64}K^4)\, .
\ee

\be
I_3=\int_{\p M_5}W_{acbd}^2=\int_{\p M_5}\left(R_{acbd}^2-\frac{16}{9}R_{ab}^2+\frac{5}{18}R^2+\frac{8}{3}R^{ab}R_{anbn}+\frac{4}{9}R_{nn}^2-\frac{8}{9}RR_{nn}\right)\, ,
\ee
\be
I_4=\int_{\p M_5}W_{anbn}^2=\int_{\p M_5}\left(\frac{1}{9}R_{ab}^2-\frac{1}{36}R^2+R_{anbn}^2-\frac{2}{3}R^{ab}R_{anbn}-\frac{4}{9}R_{nn}^2+\frac{2}{9}RR_{nn}\right)\, .
\ee
\be
\begin{split}
I_5&=\int_{\p M_5}\hat{K}^{ab}\hat{K}^{cd}W_{acbd}\\
=&\int_{\p M_5}\Big(K^{ab}K^{cd}R_{acbd}+\frac{2}{3}(K^2)^{ab}R_{ab}-\frac{5}{6}KK^{ab}R_{ab}-\frac{1}{12}\Tr K^2R+\frac{1}{8}K^2R\\
&+\frac{1}{2}KK^{ab}R_{anbn}-\frac{1}{6}K^2R_{nn}\Big)\, ,
\end{split}
\ee
\be
\begin{split}
I_6&=\int_{\p M_5}\hat{K}^a_c\hat{K}^{cb}W_{anbn}\\
=&\int_{\p M_5}\Big(-\frac{1}{3}K^a_cK^{cb}R_{ab}+\frac{1}{6}KK^{ab}R_{ab}+\frac{1}{12}\Tr K^2R-\frac{1}{24}K^2R\\
&+K^a_cK^{cb}R_{anbn}-\frac{1}{2}KK^{ab}R_{anbn}-\frac{1}{3}\Tr K^2R_{nn}+\frac{1}{6}K^2R_{nn}\Big)\, .
\end{split}
\ee
\be
\begin{split}
I_7&=\int_{\p M_5}W_{nabc}^2\\
&=\int_{\p M_5}\Big[2\nb_cK_{ab}\nb^cK^{ab}-\frac{8}{3}\nb_aK^a_b\nb_cK^{cb}+\frac{4}{3}\nb_aK\nb_bK^{ab}-\frac{2}{3}(\nb K)^2\\
&-2K^{ab}K^{cd}R_{acbd}-2(K^2)^{ab} R_{anbn}+2(K^2)^{ab} R_{ab}+2K\Tr K^3-2(\Tr K^2)^2\Big]\, .
\end{split}   
\ee
\be
\lb{I8-2}
\begin{split}
I_8&=\int_{\p M_5}\left(\Kh^{ab}\nabla_n W_{anbn}-\frac{1}{2}K\Kh^{ab}W_{anbn}-\frac{2}{9}\nb_a\Kh^a_b\nb_c\Kh^{cb}+2(\Kh^2)^{ab}\bar{S}_{ab}-\Tr \Kh^2\bar{S}^a_a\right)\\
&=\int_{\p M_5}\Big[\frac{2}{3}K^{ab}\nabla_nR_{anbn}-\frac{1}{12}K\nabla_nR+\frac{2}{3}K^{ab}K^{cd}R_{acbd}-K^a_cK^{bc}R_{ab}\\
&+\frac{1}{3}\Tr K^2 R-\frac{5}{48}K^2R+\frac{5}{3}K^a_cK^{bc}R_{anbn}-\frac{1}{3}KK^{ab}R_{anbn}-\Tr K^2R_{nn}+\frac{11}{24}K^2R_{nn}\\
&+\Tr K^4-\frac{11}{6}K\Tr K^3+\frac{47}{48}K^2\Tr K^2-\frac{7}{48}K^4\\
&-\frac{1}{3}\nb_cK_{ab}\nb^cK^{ab}+\frac{8}{9}\nb_aK^a_b\nb_cK^{bc}-\frac{7}{9}\nb_aK^{ab}\nb_bK+\frac{25}{72}(\nb K)^2\Big]\, .
\end{split}
\ee
Here we have substituted $\bar{S}_{ab}=\frac{1}{2}(\bar{R}_{ab}-\frac{1}{6}\bar{R}h_{ab})$ 
which is the  4 dimensional Schouten tensor computed with respect to the intrinsic boundary metric $h_{ab}$.

\subsubsection{Heat kernel coefficients in five dimensions}
The general form of the coefficient $B_5(x)$ in the expansion of heat kernel is given in (\ref{a5}), see \cite{Branson:1999jz}, \cite{Kirsten:2001wz}.
According to the notation of \cite{Branson:1999jz}
\be
a_5=\frac{1}{5760(4\pi)^2}\int_{\p M_5}\tr(\CA_5^1+\CA_5^2+\CA_5^3)\, ,
\ee
where $\tr$ is taken on spinor indices. 
The constituent terms  of $\tr \CA_5^{\{1,2,3\}}$ and their extended forms in terms of the curvature tensors are listed in the following three tables.
\begin{table}[H]
\small
\begin{center}
\begin{tabular}{|c c c|}
 \hline
 various terms of $\CA_5^1$ & extended forms & contribution to anomaly \\ 
  \hline
 $\nabla_n^2E$ & $-\frac{1}{4}\nabla_n^2R$ & $\tr(360\chi)=0$ \\ 
 $\tr(\nabla_nES)$ & $\frac{1}{2}K\nabla_nR$ & $-1440$ \\ 
$E^2$ & $\frac{1}{16}R^2$ & $\tr(720\chi)=0$ \\ 
$RE$ & $-\frac{1}{4}R^2$ & $\tr(240\chi)=0$ \\
$\Box R$ & $\nabla_n^2R+K\nabla_n R$ & $\tr(48\chi)=0$\\
$R^2$ & $R^2$ & $\tr(20\chi)=0$\\
$R_{ij}^2$ & $R_{ab}^2+R_{nn}^2+2R_{an}^2$ & $\tr(-8\chi)=0$\\
$R_{ikj\ell}^2$ & $R_{acbd}^2+4R_{anbn}^2+4R_{nabc}^2$ & $\tr(8\chi)=0$\\
$R_{nn}E$ & $-\frac{1}{4}RR_{nn}$ & $\tr(-120\chi)=0$\\
$RR_{nn}$ & $RR_{nn}$ & $\tr(-20\chi)=0$\\
$\tr(RS^2)$ & $K^2R$ & $480$\\
$\nabla_n^2R$ & $\nabla_n^2R$ & $\tr(12\chi)=0$\\
$\nabla_n^2R_{nn}$ & $\nabla_n^2R_{nn}$ & $\tr(15\chi)=0$\\
$\tr(\nabla_nRS)$ & $-2K\nabla_nR$ & $-270$\\
$\tr(R_{nn}S^2)$ & $K^2R_{nn}$ & $120$\\
$\tr(S\bar{\Box}S)$ & $-(\nb K)^2-\frac{1}{2}K^2\Tr K^2$ & $960$\\
$R^{ab}R_{anbn}$ & $R^{ab}R_{anbn}$ & $\tr(-16\chi)=0$\\
$R_{nn}^2$ & $R_{nn}^2$ & $\tr(-17\chi)=0$\\
$R_{anbn}^2$ & $R_{anbn}^2$ & $\tr(-10\chi)=0$\\
$\tr(ES^2)$ & $-\frac{1}{4}K^2R$ & $2880$\\
$\tr(S^4)$ & $\frac{1}{4}K^4$ & $1440$\\
\hline
\end{tabular}
\end{center}
\end{table}
\begin{table}[H]
\small
\begin{center}
\begin{longtable}{|c c c|}
 \hline
 \small
 various terms of $\CA_5^2$& extended forms & contribution to anomaly \\ 
  \hline
 $K\nabla_nE$ & $-\frac{1}{4}K\nabla_nR$ & $\tr(-90\Pi_+-450\Pi_-)=-2160$ \\ 
 $K\nabla_nR$ & $K\nabla_nR$ & $\tr(-\frac{111}{2}\Pi_+-42\Pi_-)=-390$ \\ 
 $K^{ab}\nabla_nR_{anbn}$ & $K^{ab}\nabla_nR_{anbn}$ & $\tr(-30\Pi_+)=-120$ \\ 
 $\tr(K\bar{\Box}S)$ & $2(\nb K)^2$ & $240$\\
 $\tr(K^{ab}\nb_a\nb_b S)$ & $2\nb_aK^{ab}\nb_bK$ & $420$\\
 $\tr(\nb_aK\nb^aS)$ & $-2(\nb K)^2$ & $390$\\
 $\tr(\nb_aK^{ab}\nb_bS)$ & $-2\nb_aK^{ab}\nb_bK$ & $480$\\
 $\tr(\bar{\Box}KS)$ & $2(\nb K)^2$ & $420$\\
 $\tr(\nb_a\nb_bK^{ab}S)$ & $2\nb_aK^{ab}\nb_bK$ & $60$\\
 $(\nb K)^2$ & $(\nb K)^2$ & $\tr(\frac{487}{16}\Pi_++\frac{413}{16}\Pi_-)=225$ \\ 
 $\nb_aK^{ab}\nb_bK$ & $\nb_aK^{ab}\nb_bK$ & $\tr(238\Pi_+-58\Pi_-)=720$ \\ 
 $\nb_aK^a_b\nb_cK^{bc}$ & $\nb_aK^a_b\nb_cK^{bc}$ & $\tr(\frac{49}{4}\Pi_++\frac{11}{4}\Pi_-)=60$\\
 $\nb_cK_{ab}\nb^cK^{ab}$ & $\nb_cK_{ab}\nb^cK^{ab}$ & $\tr(\frac{535}{8}\Pi_+-\frac{355}{8}\Pi_-)=90$\\
 $\nb_cK_{ab}\nb^bK^{ac}$ & $\nb_cK_{ab}\nb^bK^{ac}$ & $\tr(\frac{151}{4}\Pi_++\frac{29}{4}\Pi_-)=180$\\
$\bar{\Box}KK$ & $-(\nb K)^2$ & $\tr(111\Pi_+-6\Pi_-)=420$\\
$\nb_a\nb_bK^{ab}K$ & $-\nb_aK^{ab}\nb_b K$ & $\tr(-15\Pi_++30\Pi_-)=60$\\
 $K^{bc}\nb_c\nb^aK_{ab}$ & $-\nb_aK^a_b\nb_cK^{bc}$ & $\tr(-\frac{15}{2}\Pi_+-\frac{75}{2}\Pi_-)=120$\\
$K^{ab}\nb_a\nb_b K$ & $-\nb_aK^{ab}\nb_b K$ & $\tr(\frac{945}{4}\Pi_+-\frac{285}{4}\Pi_-)=660$\\
$K^{ab}\bar{\Box} K_{ab}$ & $-\nb_cK_{ab}\nb^cK^{ab}$ & $\tr(114\Pi_+-54\Pi_-)=240$\\
$\tr(KSE)$ & $\frac{1}{2}K^2R$ & $1440$\\
$\tr(KSR_{nn})$ & $-2K^2R_{nn}$ & $30$\\
$\tr(KSR)$ & $-2K^2R$ & $240$\\
$\tr(K^{ab}R_{ab}S)$ & $-2KK^{ab}R_{ab}$ & $-60$\\
$\tr(K^{ab}SR_{anbn})$ & $-2KK^{ab}R_{anbn}$ & $180$\\
$K^2E$ & $-\frac{1}{4}K^2R$ & $\tr(195\Pi_+-105\Pi_-)=360$\\
$\Tr K^2E$ & $-\frac{1}{4}\Tr K^2R$ & $\tr(30\Pi_++150\Pi_-)=720$\\
$K^2R$ & $K^2R$ & $\tr(\frac{195}{6}\Pi_+-\frac{105}{6}\Pi_-)=60$\\
$\Tr K^2R$ & $\Tr K^2R$ & $\tr(5\Pi_++25\Pi_-)=120$\\
$K^2R_{nn}$ & $K^2R_{nn}$ & $\tr(-\frac{275}{16}\Pi_++\frac{215}{16}\Pi_-)=-15$\\
$\Tr K^2R_{nn}$ & $\Tr K^2R_{nn}$ & $\tr(-\frac{275}{8}\Pi_++\frac{215}{8}\Pi_-)=-30$\\
$KK^{ab}R_{ab}$ & $KK^{ab}R_{ab}$ & $\tr(-\Pi_+-14\Pi_-)=-60$\\
$KK^{ab}R_{anbn}$ & $KK^{ab}R_{anbn}$ & $\tr(\frac{109}{4}\Pi_+-\frac{49}{4}\Pi_-)=60$\\
$K^a_cK^{bc}R_{ab}$ & $K^a_cK^{bc}R_{ab}$ & $\tr(16\Pi_+-16\Pi_-)=0$\\
$K^a_cK^{bc}R_{anbn}$ & $K^a_cK^{bc}R_{anbn}$ & $\tr(\frac{133}{2}\Pi_++\frac{47}{2}\Pi_-)=360$\\
$K^{ab}K^{cd}R_{acbd}$ & $K^{ab}K^{cd}R_{acbd}$ & $\tr(32\Pi_+-32\Pi_-)=0$\\
$\tr(KS^3)$ & $-\frac{1}{2}K^4$ & $2160$\\
$K^2S^2$ & $K^4$ & $1080$\\
$\Tr K^2S^2$ & $K^2\Tr K^2$ & $360$\\
$K^3S$ & $-2K^4$ & $\frac{885}{4}$\\
$K\Tr K^2S$ & $-2K^2\Tr K^2$ & $\frac{315}{2}$\\
$\Tr K^3S$ & $-2K\Tr K^3$ & $150$\\
$K^4$ & $K^4$ & $\tr(\frac{2041}{128}\Pi_++\frac{65}{128}\Pi_-)=\frac{1053}{16}$\\
$K^2\Tr K^2$ & $K^2\Tr K^2$ & $\tr(\frac{417}{32}\Pi_++\frac{141}{32}\Pi_-)=\frac{279}{4}$\\
$(\Tr K^2)^2$ & $(\Tr K^2)^2$ & $\tr(\frac{375}{32}\Pi_+-\frac{777}{32}\Pi_-)=-\frac{201}{4}$\\
$K\Tr K^3$ & $K\Tr K^3$ & $\tr(25\Pi_+-\frac{17}{2}\Pi_-)=66$\\
$\Tr K^4$ & $\Tr K^4$ & $\tr(\frac{231}{8}\Pi_++\frac{327}{8}\Pi_-)=279$\\
\hline
\end{longtable}
\end{center}  
\end{table}
\begin{table}[H]
\small
\begin{center}
\begin{tabular}{|c c c|}
 \hline
 various terms of $\CA_5^3$ & extended forms & contribution to anomaly \\ 
  \hline
$E^2$ & $\frac{1}{16}R^2$ & $\tr(-180\mathbb{I})=-1440$ \\ 
$\chi E\chi E$ & $\frac{1}{16}R^2$ & $\tr(180\chi^2)=1440$ \\ 
$\tr((\nb S)^2)$ & $(\nb K)^2+\frac{1}{2}K^2\Tr K^2$ & $-120$\\
$\tr(\chi(\nb S)^2)$ & $(\nb K)^2$ & $720$\\
$\tr(\Omega_{ab}\Omega^{ab})$ & $-R_{acbd}^2-2R_{nabc}^2$ & $-\frac{105}{4}$\\
$\tr(\chi\Omega_{ab}\Omega^{ab})$ & $0$ & $120$\\
$\tr(\chi\Omega_{ab}\chi\Omega^{ab})$ & $-R_{acbd}^2+2R_{nabc}^2$ & $\frac{105}{4}$\\
$\tr(\Omega_{an}\Omega^a_n)$ & $-2R_{anbn}^2-R_{nabc}^2$ & $-45$\\
$\tr(\chi\Omega_{an}\Omega^a_n)$ & $0$ & $180$\\
$\tr(\chi\Omega_{an}\chi\Omega^a_n)$ & $2R_{anbn}^2-R_{nabc}^2$ & $-45$\\
$\tr(\Omega_{an}\chi\nb^aS-\Omega_{an}\nb^aS\chi)$ & $-2KK^{ab}R_{anbn}$ & $-360$\\
$\tr(\chi\nb_a\chi\Omega^a_nK)$ & $4KK^{ab}R_{anbn}$ & $-45$\\
$\tr(\nb_a\chi\nb_b\chi\Omega^{ab})$ & $-4K^{ab}K^{cd}R_{acbd}$ & $-180$\\
$\tr(\chi\nb_a\chi\nb_b\chi\Omega^{ab})$ & $0$ & $30$\\
$\tr(\chi\nb_a\chi\nabla_n\Omega^a_n)$ & $4K^{ab}\nabla_n R_{anbn}$ & $90$\\
$\tr(\chi\nb^a\chi\nb^b\Omega_{ab})$ & $-4K^{ab}K^{cd}R_{acbd}-4\nb^bK^{ac}R_{abcn}$ & $120$\\
$\tr(\chi\nb_a\chi\Omega_{bn}K^{ab})$ & $4K^a_cK^{bc}R_{anbn}$ & $-180$\\
$\tr(\nb_a\chi\nb^aE)$ & $0$ & $300$\\
$\tr(\nb_a\chi\nb^a\chi E)$ & $-2\Tr K^2 R$ & $-180$\\
$\tr(\chi\nb_a\chi\nb^a\chi E)$ & $0$ & $-90$\\
$\tr(\bar{\Box}\chi E)$ & $0$ & $240$\\
$\tr(\nb_a\chi\nb^a\chi R)$ & $8\Tr K^2 R$ & $-30$\\
$\tr(\nb_a\chi\nb_b\chi R^{ab})$ & $8K^a_cK^{bc}R_{ab} $ & $-60$\\
$\tr(\nb_a\chi\nb_b\chi R_n\,^{ab}\,_n)$ & $-8K^a_cK^{bc}R_{anbn} $ & $-30$\\
$\tr(\nb_a\chi\nb^a\chi K^2)$ & $8K^2\Tr K^2$ & $-\frac{675}{32}$\\
$\tr(\nb_a\chi\nb_b\chi K^a_cK^{bc})$ & $8\Tr K^4$ & $-\frac{75}{4}$\\
$\tr(\nb_a\chi\nb^a\chi \Tr K^2)$ & $8(\Tr K^2)^2$ & $-\frac{195}{16}$\\
$\tr(\nb_a\chi\nb_b\chi K^{ab}K)$ & $8K\Tr K^3$ & $-\frac{675}{8}$\\
$\tr(\nb_a\chi\nb^a S K)$ & $-2K^2\Tr K^2$ & $-330$\\
$\tr(\nb_a\chi\nb_b S K^{ab})$ & $-2K\Tr K^3$ & $-300$\\
$\tr(\nb_a\chi\nb^a\chi\nb_b\chi\nb^b\chi)$ & $8(\Tr K^2)^2$ & $\frac{15}{4}$\\
$\tr(\nb_a\chi\nb_b\chi\nb^a\chi\nb^b\chi)$ & $16\Tr K^4-8(\Tr K^2)^2$ & $\frac{15}{8}$\\
$\tr(\bar{\Box}\chi\bar{\Box}\chi)$ & $8\nb_aK^a_b\nb_cK^{bc}+8(\Tr K^2)^2$ & $-\frac{15}{4}$\\
$\tr(\nb_a\nb_b\chi\nb^a\nb^b\chi)$ & $8\nb_cK_{ab}\nb^cK^{ab}+8\Tr K^4$ & $-\frac{105}{2}$\\
$\tr(\nb_a\chi\nb^a\chi\bar{\Box}\chi)$ & $0$ & $-15$\\
$\tr(\nb_a\bar{\Box}\chi\nb^a\chi)$ & $-8\nb_aK^a_b\nb_cK^{bc}-8(\Tr K^2)^2$ & $-\frac{135}{2}$\\
\hline
\end{tabular}
\end{center}
\end{table}
In derivation of the second columns of the tables we have extensively used the identities in \eqref{App1} and that $\tr \Pi_\pm=4$ in the doubling representation in $d=5$.
By adding the similar terms of the second columns, taking into account the coefficients of the third columns and after using the identities of section \eqref{Gauss-Codazzi}, we get the following result
\be
\begin{split}
a_5&=\frac{1}{5760(4\pi)^2}\int_{\p\CM_5}\Big(240K^{ab}\nabla_n R_{anbn}-30K\nabla_nR\\ 
&-450K^{ab}K^{cd}R_{acbd}+210K^a_cK^{bc}R_{ab}+60KK^{ab}R_{ab}+60\Tr K^2R-15K^2R\\
&-810K^a_cK^{bc}R_{anbn}+240KK^{ab}R_{anbn}-30\Tr K^2 R_{nn}+45K^2R_{nn}\\
&-261\Tr K^4+381K\Tr K^3-\frac{1251}{4}(\Tr K^2)^2+66K^2\Tr K^2-\frac{267}{16}K^4\\
&+300\nb_cK_{ab}\nb^cK^{ab}-240\nb_aK^a_b\nb_cK^{bc}-15(\nb K)^2\Big)\,  .
\end{split}
\ee
It is a curious observation that in this expression the Riemann curvature appears only in combination with the extrinsic curvature. 
The anomaly thus vanishes for a geodesic boundary.
A technical explanation for this fact
is that in the heat kernel (\ref{a5}) all terms that are expressed only in terms of the Riemann curvature come with matrix $\chi=\Pi_+-\Pi_-$ whose trace vanishes.
In the case of a scalar field considered in \cite{FarajiAstaneh:2021foi}  one has that $\chi$ is either $+1$ or $-1$ depending on the type of the boundary condition.
As a result, the curvature terms are present in $a_5$ and in the anomaly.
For exactly same reason such terms did not appear in $d=3$ dimensions for the Dirac field and appeared for the conformal scalars.

\subsubsection{Conformal charges in $d=5$ }

This expression gives us the integrated conformal anomaly
\be
\int_{\p M_5}\la T_{ij}\ra g^{ij}=-a_5\, .
\lb{anomaly5d}
\ee
Now it has to be presented as a combination of the conformal invariants, $E_4$ and $I_1$ to $I_8$, and compared to the general form of the anomaly (\ref{anomaly5}) and determine the conformal charges $(a, c_1,\dots , c_8)$. Doing this we arrive at 27 algebraic equations for 9 conformal charges that we present in appendix \eqref{appE}.
These equations have unique solution that gives the following values for the conformal charges:
\begin{table}[H]
\begin{center}
\renewcommand{\baselinestretch}{2}
\medskip
\bigskip
 \begin{tabular}[H]{ | c | c | }
    \hline
  \text{Conformal charges} & \text{Mixed b.c.}\\
  \hline
  $a$ & $0$ \\
  \hline
$c_1$ & $-\frac{429}{4}$\\
  \hline
$c_2$ & $621$  \\
  \hline
 $c_3$ & $0$ \\
  \hline
$c_4$ & $0$  \\
  \hline
$c_5$ & $270$ \\
  \hline
$c_6$ & $990$  \\
  \hline
$c_7$ & $-210$ \\
  \hline
$c_8$ & $-360$\\
  \hline
\end{tabular}
\renewcommand{\baselinestretch}{1}
\end{center}
\end{table}

\subsubsection{Comparison with conformal scalar field}
It is instructive to compare the found charges for the fermions with the charges computed earlier in \cite{FarajiAstaneh:2021foi}
for a conformal scalar field satisfying either Dirichlet boundary condition or conformal Robin boundary condition.
\begin{table}[H]
\begin{center}
\renewcommand{\baselinestretch}{2}
\medskip
\bigskip
 \begin{tabular}{ | c | c | c | }
    \hline
  \text{Conformal charges} & \text{Dirichlet b.c.} & \text{Robin b.c.} \\
  \hline
  $a$ & $\frac{17}{8}$ & $-\frac{17}{8}$ \\
  \hline
$c_1$ & $-\frac{681}{32}$ & $\frac{39}{32}$ \\
  \hline
$c_2$ & $\frac{609}{8}$ & $\frac{309}{8}$ \\
  \hline
 $c_3$ & $-\frac{81}{8}$ & $\frac{81}{8}$ \\
  \hline
$c_4$ & $-\frac{27}{2}$ & $\frac{27}{2}$ \\
  \hline
$c_5$ & $-\frac{9}{8}$ & $\frac{189}{8}$ \\
  \hline
$c_6$ & $\frac{819}{8}$ & $\frac{441}{8}$ \\
  \hline
$c_7$ & $-\frac{615}{16}$ & $\frac{195}{16}$ \\
  \hline
$c_8$ & $-\frac{45}{2}$ & $-\frac{45}{2}$ \\
  \hline
\end{tabular}
\renewcommand{\baselinestretch}{1}
\end{center}
\end{table}
Denoting the charges for Dirichlet, Robin and Mixed boundary conditions with indices, $D$, $R$ and $M$ respectively, the following relations are observed between some charges.
For invariants which are written purely in terms of the Riemann tensor and its contractions we find  a relation
\be
a^M=a^D+a^R=0 \ \ , \ \ c^M_{3,4}=c^D_{3,4}+c^R_{3,4}=0\, .
\ee
For the charge $a$ we saw a similar relation in $d=3$ dimensions.
For invariants which include the derivatives, we find a relation
\be
c^M_{7,8}=8(c^D_{7,8}+c^R_{7,8})\, .
\ee
No obvious relations were observed for the other charges.

\subsubsection{Including the gauge fields}
Let us now discuss the possible contribution of the gauge fields to the conformal anomaly in five dimensions.
There are two possible terms in the anomaly that we present in the form,
\be
 \frac{1}{1536\pi^2} \int_{\p M_5}(b_1 F_{ab} F^{ab}+b_2 F_{an}F^{a}_{\ n})\, ,
\lb{anomalygauge}
\ee
that are due to the gauge fields. In the heat kernel coefficient $a_5$ the gauge field $A_i$ with the field strength $F_{ij}$
may appear either via $E=-\frac{1}{4}R+\frac{1}{2}F_{ij}\gamma^i\gamma^j$ or via $\Omega_{ij}=\frac{1}{4}R_{ijk\ell}\gamma^k\gamma^\ell+F_{ij}$.
There is a plenty of such terms in the heat kernel coefficient (\ref{a5}). Most of them give zero after taking the trace over spinor indexes. The non-vanishing terms are given in the table below.
\begin{table}[H]
\begin{center}
\begin{tabular}{|c c c|}
 \hline
 Various terms including $F_{ij}$ & extended forms & contribution to anomaly \\ 
  \hline
$\tr(E^2)$ & $-4F_{ab}F^{ab}-8F_{an}F^{an}$ & $-180$\\
$\tr(\chi E\chi E)$ & $-4F_{ab}F^{ab}+8F_{an}F^{an}$ & $180$\\
$\tr(\Omega_{ab}\Omega^{ab})$ & $8F_{ab}F^{ab}$ & $-\frac{105}{4}$\\
$\tr(\chi \Omega_{ab}\chi \Omega^{ab})$ & $8F_{ab}F^{ab}$ & $\frac{105}{4}$\\
$\tr(\Omega_{an}\Omega^a_n)$ & $8F_{an}F^a_n$ & $-45$\\
$\tr(\chi \Omega_{an}\chi \Omega^a_n)$ & $8F_{an}F^a_n$ & $-45$\\
\hline
\end{tabular}
\end{center}
\end{table}
\noindent We have also checked that the possible cross terms that contain both $F_{ij}$ and the Riemann curvature do not appear.

Collecting all terms together we find that
the anomaly
\be
\int_{\p M_5}\la T_{ij}\ra g^{ij}=-\frac{3}{128\pi^2}\int_{\p M_5} F_{an}F^{a}_{\ n}\, .
\ee
Comparing with (\ref{anomalygauge}) we conclude that in the anomaly  for a Dirac fermion the charge $b_1=0$ while the only non-vanishing charge is $b_2$.
At the moment we do not have an explanation for this result.

This result is worth comparing with the gauge field terms in the conformal anomaly for a complex scalar field carrying a representation for the gauge group $G$ and coupled to the gauge fields. 
In this case $\chi=-1$ for the Dirichlet boundary conditions and 
$\chi=+1$ for the conformal Robin boundary conditions, see   \cite{FarajiAstaneh:2021foi}. We have computed these terms. Only two terms in the heat kernel coefficient contribute to the anomaly in this case: $\Omega_{ab}\Omega^{ab}$ and $\Omega_{an}\Omega^{an}$. Omitting the details that are quite simple the result is summarised below.

\begin{table}[H]
\begin{center}
\renewcommand{\baselinestretch}{2}
\medskip
\bigskip
 \begin{tabular}[H]{ | c | c | c | c | }
    \hline
  \text{Conformal charges} & \text{Dirichlet b.c.} & \text{Robin b.c.} & \text{Mixed b.c.}\\
  \hline
$b_1$ & $4$ & $9$ & $0$\\
  \hline
$b_2$ & $-4$ & $-3$ & $36$  \\
  \hline 
\end{tabular}
\renewcommand{\baselinestretch}{1}
\end{center}
\end{table}
We find a relation
\be
b^M_k=6(b^D_k+b^R_k)\, , \  k=1,2\, 
\ee
between the charges for fermions (M) and scalars with Dirichlet (D) and Robin (R) boundary conditions.

\subsection{Chiral anomaly in $d=5$ dimensions}
\subsubsection{Parity odd conformal invariants}
There are three parity odd conformal  invariants in five dimensions, see \cite{Chalabi:2021jud},
\be
J_1=\int_{\p M_5}\epsilon^{abcd}W_{abef}W_{cd}\,^{ef}\, ,
\ee
\be
J_2=\int_{\p M_5}\epsilon^{abcd}W_{abne}W_{cdn}\,^{e}=4\int_{\p M_4}\epsilon^{abcd}\nb_b\Kh_{ae}\nb_d\Kh^e_c\, ,
\ee
and
\be
J_3=\int_{\p M_5}\epsilon^{abcd}\Kh^e_a\Kh^f_bW_{cdef}\, ,
\ee
where $W_{abcd}$ is the bulk Weyl tensor with the boundary indices and $\Kh_{ab}$ is the traceless part of the extrinsic curvature tensor.
If we include the gauge fields then there is one more invariant\,
\be
J_4=\int_{\p M_5}\epsilon^{nabcd}F_{ab}F_{cd}\, .
\ee
The chiral anomaly decomposes as follows,
\be\label{gch}
\int_{M_4}\nabla_i \la j^i_{A} \ra= -\frac{1}{96(4\pi)^2}(d_1 J_1+d_2 J_2+d_3 J_3+ d_4 J_4)\, .
\ee

\subsubsection{Computation using heat kernel coefficients}
For the Dirac field the anomaly is 
\be
\int_{M_4}\nabla_i \la j^i_{A} \ra=-2\int_{\p M_5}\tr (i\Gamma^*_2 B_5(x))\, .
\lb{ch-5}
\ee
There are four terms\footnote{The coefficient of $\chi\nb_a\chi\nb_b\chi\Omega^{ab}$ was reported as $90$ for the first time in \cite{Branson:1999jz}. Later in \cite{Moss:2012dp}, I. Moss corrected this and reported this coefficient as $30$. We thank the referee for bringing this to our attention.} in the heat kernel that produce a non-trivial trace in (\ref{ch-5}).
\begin{table}[H]
\begin{center}
\begin{tabular}{|c c c|}
 \hline
 various parity odd terms & extended forms & contribution to anomaly \\ 
  \hline
$\tr(i\Gamma^*_2\chi\Omega_{ab}\Omega^{ab})$ & $\frac{1}{2}\epsilon^{nabcd}R_{abef}R_{cd}\,^{ef}$ & $120$\\
$\tr(i\Gamma^*_2\chi\Omega_{an}\Omega^a_n)$ & $\frac{1}{2}\epsilon^{nabcd}R_{abne}R_{cdn}\,^{e}=2\epsilon^{nabcd}\nb_bK_{ae}\nb_dK^e_c$ & $180$\\
$\tr(i\Gamma^*_2\chi\nb_a\chi\nb_b\chi\Omega^{ab})$ & $2\epsilon^{nabcd}K^e_aK^f_bR_{cdef}$ & $30$\\
$\tr(i\Gamma^*_2\chi E^2)$ & $2\epsilon^{nabcd}F_{ab}F_{cd}$ & $720$\\
\hline
\end{tabular}
\end{center}
\end{table}
It is not difficult to see that in these invariants Riemann tensor can be replaced by Weyl tensor and the extrinsic curvature by its trace-free part so that these are precisely invariants
that we listed above.
Matching the coefficients with the general form (\ref{gch})  we find the following corresponding charges.
\begin{table}[H]
\begin{center}
\renewcommand{\baselinestretch}{2}
\medskip
\bigskip
 \begin{tabular}[H]{ | c | c | }
    \hline
  \text{Conformal charges} & \text{Mixed b.c.}\\
  \hline
$d_1$ & $2$\\
  \hline
$d_2$ & $3$  \\
  \hline
 $d_3$ & $2$ \\
  \hline
  $d_4$ & $48$ \\
  \hline 
\end{tabular}
\renewcommand{\baselinestretch}{1}
\end{center}
\end{table}

\bigskip

This completes our consideration of anomaly in $d=5$ dimensional spacetime.

\section{Conclusions}
That the quantum anomalies are  modified in the presence of boundaries by the boundary terms 
is an interesting subject of research that came into light in the recent years.
In the present paper we have developed a systematic calculation for the boundary terms in the conformal anomaly and
in the chiral anomaly. The conformal anomaly in even dimensional spacetime in this context was studied in
\cite{Fursaev:2015wpa}, \cite{Herzog:2015ioa} (see also earlier paper \cite{Dowker:1989ue}) where the anomaly in $d=4$ was systematically studied.
It is intriguing that the both anomalies which are usually absent in odd dimensions, can be non-trivial in the presence of boundaries. For the conformal anomaly this is known already for some time \cite{Solodukhin:2015eca}.  The complete basis of conformal boundary terms
in the anomaly in $d=5$ was identified in \cite{FarajiAstaneh:2021foi} where the respective conformal charges for a scalar field with either Dirichlet or Robin boundary conditions were
computed.  In the present paper we continued the previous study in $d=5$ and have computed the conformal charges for Dirac fermions in $d=5$ dimensions.
A new subject of research that is in the focus of the present paper is the boundary terms in the chiral anomaly. To the best of our knowledge, this issue was not  widely discussed 
before. An earlier paper, known to us,  on this subject is \cite{Marachevsky:2003zb} where in $d=4$ dimensions, the boundary terms in chiral anomaly due to the axial gauge fields (not considered in the present paper) were found
in a rather restrictive case when the boundary is geodesic. We should  also mention here the papers by Vassilevich et al. \cite{Kurkov:2018pjw} - \cite{Fresneda:2023wub} on parity anomaly. This anomaly  is different from the chiral anomaly although it has a  similar odd  structure.

\vspace{0.5cm}
\noindent Below we summarise our findings:
\vspace{0.3cm}

\noindent $\bullet$ Boundary terms due to gauge field in chiral anomaly in $d=3$ dimensions.

\vspace{0.3cm}
\noindent $\bullet$ Gravitational boundary term in chiral anomaly in $d=4$ dimensions.

\vspace{0.3cm}
\noindent $\bullet$ Boundary conformal anomaly for fermions in dimension $d=5$ both due to the gravitational field and the gauge fields. 

\vspace{0.3cm}
\noindent $\bullet$ The anomaly due to
gauge fields for conformal scalars with either Dirichlet or Robin boundary condition   in $d=5$ dimensions that completes our previous study \cite{FarajiAstaneh:2021foi}
of anomaly for conformal scalars.

\vspace{0.3cm} 
\noindent $\bullet$ Boundary terms in chiral anomaly in $d=5$ dimensions both due to the gravitational field and the gauge fields.

\vspace{0.5cm}
It would be interesting to develop the holographic aspects for the present calculations of the anomaly, this is the subject of a work in progress \cite{WIP}. It would also be interesting to find some applications for our findings such as the chiral anomaly in $d=3$ or the boundary term in the chiral anomaly in $d=4$.
We leave these issues for a further study.

\newpage
\appendix
\section{Identities for matrices and traces}\label{App1}
\setcounter{equation}0
\numberwithin{equation}{section}
\be
\gamma^n\Pi_\pm=\Pi_\mp\gamma^n \ \ , \ \  \gamma^a\Pi_\pm=\Pi_\pm\gamma^a \ \ , \ \ \chi\Pi_\pm=\Pi_\pm\, .
\ee
\be\label{I2}
\nabla_a\gamma^*=0 \ \ , \ \ \nabla_a\gamma^n=K_{ab}\gamma^b\ \ , \ \ \nabla_a\gamma^b=-K_a^b\gamma^n\, .
\ee
\be\label{A3}
\begin{split}
&\nb_a\chi=i\gamma^*\gamma^bK_{ab}\\
&\nb_a\nb_b\chi=i\gamma^*(\gamma^c\nb_aK_{bc}+\gamma^nK^2_{ab})\, ,\\
\end{split}
\ee
We also need the intrinsic derivatives of $S=-\frac{1}{2}K\Pi_+$, which are calculated below
\be
\begin{split}
&\nb_aS=-\frac{1}{2}\nb_aK\Pi_+-\frac{1}{4}K\nb_a\chi\, ,\\
&\nb_a\nb_bS=-\frac{1}{2}(\nb_a\nb_bK)\Pi_+-\frac{1}{4}K\nb_a\nb_b\chi-\frac{1}{4}(\nb_aK\nb_b\chi+\nb_bK\nb_a\chi)\, .
\end{split}
\ee
These identities rely only on the commutation relations and do not depend on the choice of the representation for gamma matrices and for the chirality matrix $\gamma^*$.

\section{Identities}
\underline{\textbf{Gauss-Codazzi relations}}\label{Gauss-Codazzi}
\be\label{GC1}
R_{acbd}=\bar{R}_{acbd}-(K_{ab}K_{cd}-K_{ad}K_{bc})\, ,
\ee
\be\label{GC2}
R_{nabc}=(\bar{\nabla}_cK_{ab}-\bar{\nabla}_bK_{ac})\, ,
\ee
where $\bar{R}_{acbd}$ represents the intrinsic Riemann tensor of the boundary. 
In particular, in five dimensions we need
\be
R_{nabc}^2=2\nb_cK_{ab}\nb^cK^{ab}-2\nb_cK_{ab}\nb^bK^{ac}\, ,
\ee
where the second term can be expanded as \eqref{identity 4}.

The contracted equations read
\be\label{CGC1}
R_{an}=R_{na}=(\bar{\nabla}_bK^b_a-\bar{\nabla}_aK)\, .
\ee
\be\label{CGC2}
R_{ab}=\bar{R}_{ab}+ R_{anbn}+(K^2_{ab}-KK_{ab})\, ,
\ee
and a double contraction yields
\be
R=\bar{R}+2 R_{nn}+(\Tr K^2-K^2)\, .
\ee
Thus, one finds for the projected Einstein tensor
\be
G_{nn}=-\frac{1}{2}\bar{R}-\frac{1}{2}(\Tr K^2-K^2)\, .
\ee
\\
\underline{\textbf{Differential Equations}} 
\be\label{identity 1}
\Box R=\bar{\Box}R+\nabla_n^2R+K\nabla_n R\, ,
\ee
\be\label{identity 2}
\nabla_nG_{nn}=K^{ab}R_{ab}-KR_{nn}-\bar{\nabla}_a\bar{\nabla}_bK^{ab}+\bar{\Box}K\, ,
\ee
\be\label{identity 4}
\begin{split}
\nb_c K_{ab}\nb^b K^{ac}=\nb_aK^a_b\nb_cK^{bc}+K^{ab}K^{cd}R_{acbd}-K^{ac}K^b_cR_{ab}\\
+K^{ac}K^b_cR_{anbn}-K\Tr K^3+(\Tr K^2)^2+T.D.\, ,
\end{split}
\ee
\be\label{identity 5}
\begin{split}
 \nabla_n^2G_{nn}&=-R^{ab}R_{anbn}+R_{nn}^2
-K^a_cK^{bc}R_{ab}+\Tr K^2R_{nn}\\
&+K^{ab}\nabla_nR_{ab}-K\nabla_n R_{nn}-\nb_aK^{ab}\nb_bK+(\nb K)^2+T.D.\, ,
\end{split}
\ee
\be\label{identity 3}
\begin{split}
\nabla_nR_{ab}&=\nabla_nR_{anbn}-2K^{cd}R_{acbd}-K^c_aR_{bncn}-K^c_bR_{ancn}+KR_{anbn}+K_{ab}R_{nn}\\
&-\Tr K^2K_{ab}+KK_{ac}K^c_b+\nb_a\nb_cK^c_b+\nb_b\nb_cK^c_a-\bar{\Box}K_{ab}-\nb_a\nb_bK\, ,
\end{split}
\ee
where we defined $\nabla_n G_{nn}=n^k n^i n^j \nabla_k G_{ij}$, $\Box G_{nn}=n^in^j\Box G_{ij}$ and  $ \nabla_n^2 G_{nn}=n^k n^\ell n^i n^j \nabla_k\nabla_\ell G_{ij}$.

\section{Derivation of the Robin boundary condition}\label{appC}
Since Dirac operator is a first order operator, one imposes the Dirichlet  boundary condition on a half of the Dirac spinor components
\be
\Pi_-\psi\vert_{\p M}=\frac{1}{2}(1-i\gamma^*\gamma^n)\psi\vert_{\p M}=0\, .
\ee
Acting the projector on both sides of the eigenvalue equation, $\gamma^k\nabla_k\psi=\lambda\psi$ one gets zero again, thus
\be
\Pi_-(\gamma^k\nabla_k)\psi\vert_{\p M}=0\, .
\ee
Putting $\gamma^n\Pi_-$ on the left hand side, we get
\be
\Pi_+\gamma^n\Pi_-(\gamma^k\nabla_k)\psi\vert_{\p M}=0\, ,
\ee
where we have defined $\Pi_+=\frac{1}{2}(1+i\gamma^*\gamma^n)$ and have used
\be
\gamma^n\Pi_-=\Pi_+\gamma^n\, .
\ee
By separating the normal and tangential components, we will have
\be
\begin{split}
&\Pi_+\gamma^n\Pi_-(\gamma^n\nabla_n+\gamma^a\nabla_a)\psi\vert_{\p M}=0\\
&\rightarrow\Pi_+^2\gamma^n\nabla_n\psi\vert_{\p M}+\Pi_+\gamma^n\gamma^a\Pi_-\nabla_a\psi\vert_{\p M}=0\\
&\rightarrow \Pi_+\nabla_n\psi\vert_{\p M}-\Pi_+\gamma^n\gamma^a(\nabla_a\Pi_-)\psi\vert_{\p M}=0\\
&\rightarrow \Pi_+(\nabla_n+\frac{1}{2}\gamma^n\gamma^a\nabla_a\chi)\psi\vert_{\p M}=0\\
&\rightarrow (\nabla_n+\frac{1}{2}\Pi_+\gamma^n\gamma^a\nabla_a\chi)\Pi_+\psi\vert_{\p M}=0\, ,
\end{split}
\ee
where we have defined $\chi=\Pi_+-\Pi_-=i\gamma^*\gamma^n$. Comparing with $(\nabla_n-S)\Pi_+\psi\vert_{\p M}=0$ we conclude that
\be
S=-\frac{1}{2}\Pi_+\gamma^n\gamma^a\nabla_a\chi\Pi_+\, .
\ee
Now using \eqref{I2} and \eqref{A3}
one deduces
\be
S=-\frac{1}{2}K\Pi_+\, .
\ee

\section{Bulk and boundary terms in the heat kernel coefficient $a_d$  in dimension $d$ \cite{Vassilevich:2003xt}, \cite{Branson:1999jz}}

\subsection{Heat kernel coefficient in $d=3$}
\be
\begin{split}
B_3(x)&=\frac{1}{384(4\pi)}\Big[96\chi E+16\chi R-8\chi R_{nn}+(2\Pi_++10\Pi_-)\Tr K^2+(13\Pi_+-7\Pi_-)K^2\\
&+96 SK+192S^2-12\nb_a\chi\nb^a\chi\Big]\, .
\end{split}
\lb{a3}
\ee
\subsection{Heat kernel coefficient in $d=4$ }
\be
\begin{split}
A_4(x)&=\frac{1}{360(4\pi)^2}\Big(60\Box E+12\Box R+2R_{ikj\ell}R^{ikj\ell}-2R_{ij}R^{ij}+180E^2+60RE+5R^2+30\Omega_{ij}\Omega^{ij}\Big)\\
B_4(x)&=\frac{1}{360(4\pi)^2}\Big[(-240\Pi_++120\Pi_-)\nabla_nE+(-42\Pi_++18\Pi_-)\nabla_nR+24\bar{\Box}K\\
&+120KE+20KR-4KR_{nn}+16K^{ab}R_{anbn}-4K^{ab}R_{ab}\\
&+\frac{1}{21}(224\Pi_++320\Pi_-)\Tr K^3+\frac{1}{21}(168\Pi_+-264\Pi_-)K\Tr K^2+\frac{1}{21}(280\Pi_++40\Pi_-)K^3\\
&+720SE+120SR+48S\Tr K^2+144SK^2+480S^2K+480S^3+120\bar{\Box}S\\
&-60\chi\nb^a\chi\Omega_{an}-24K^{ab}\nb_a\chi\nb_b\chi-12K\nb_a\chi\nb^a\chi-120S\nb_a\chi\nb^a\chi\Big]\, .
\lb{a4}
\end{split}
\ee
\subsection{Heat kernel coefficient in $d=5$ }
\be
\begin{split}
B_5(x)&=\frac{1}{5760(4\pi)^2}\Big[360\chi\nabla_n^2E-1440\nabla_nES+720\chi E^2+240\chi\bar{\Box}E+240\chi RE+48\chi\nabla^2 R+20\chi R^2\\
&-8\chi R^2_{ij}+8\chi R^2_{ikj\ell}-120\chi R_{nn}E-20\chi RR_{nn}+480RS^2+12\chi\nabla_n^2R\\
&+24\chi\bar{\Box}R_{nn}+15\chi\nabla_n^2R_{nn}-270S\nabla_nR+120R_{nn}S^2+960S\bar{\Box}S+16\chi R^{ab}R_{anbn}\\
&-17\chi R_{nn}^2-10\chi R_{anbn}R^a\,_n\,^b\,_n+2880ES^2+1440S^4\\
&-(90\Pi_++450\Pi_-)K\nabla_n E-(\frac{111}{2}\Pi_++42\Pi_-)K\nabla_nR-30\Pi_+K^{ab}\nabla_nR_{anbn}+240K\bar{\Box}S\\
&+420K^{ab}\bar{\nabla}_a\bar{\nabla}_bS+390\bar{\nabla}_aK\bar{\nabla}^aS+480\bar{\nabla}_aK^{ab}\bar{\nabla}_bS +420S\bar{\Box}K+60\bar{\nabla}_a\bar{\nabla}_bK^{ab}S\\
&+(\frac{487}{16}\Pi_++\frac{413}{16}\Pi_-)(\bar{\nabla}K)^2+(238\Pi_+-58\Pi_-)\bar{\nabla}_aK^{ab}\bar{\nabla}_bK+(\frac{49}{4}\Pi_++\frac{11}{4}\Pi_-)\bar{\nabla}_aK^{ab}\bar{\nabla}_cK^c_b\\
&+(\frac{535}{8}\Pi_+-\frac{355}{8}\Pi_-)\bar{\nabla}_cK^{ab}\bar{\nabla}^cK_{ab} +(\frac{151}{4}\Pi_++\frac{29}{4}\Pi_-)\bar{\nabla}_cK_{ab}\bar{\nabla}^bK^{ac}+(111\Pi_+-6\Pi_-)K\bar{\Box}K\\ &+(-15\Pi_++30\Pi_-)K\bar{\nabla}_a\bar{\nabla}_bK^{ab}+(-\frac{15}{2}\Pi_++\frac{75}{2}\Pi_-)K^{bc}\bar{\nabla}_c\bar{\nabla}_aK^a_b+(\frac{945}{4}\Pi_+-\frac{285}{4}\Pi_-)K^{ab}\bar{\nabla}_a\bar{\nabla}_bK\\
&+(114\Pi_+-54\Pi_-)K^{ab}\bar{\Box}K_{ab}+1440KSE+30KSR_{nn}+240KSR-60K_{ab}R^{ab}S+180K^{ab}R_{anbn}S\\ &+(195\Pi_+-105\Pi_-)K^2E+(30\Pi_++150\Pi_-)\Tr K^2E+(\frac{195}{6}\Pi_+-\frac{105}{6}\Pi_-)K^2R\\
&+(5\Pi_++25\Pi_-)\Tr K^2R+(-\frac{275}{16}\Pi_++\frac{215}{16}\Pi_-)K^2R_{nn}+(-\frac{275}{8}\Pi_++\frac{215}{8}\Pi_-)\Tr K^2R_{nn}\\
&+(-\Pi_+-14\Pi_-)KK^{ab}R_{ab}+(\frac{109}{4}\Pi_+-\frac{49}{4}\Pi_-)KK^{ab}R_{anbn}+16\chi K_{ab}K^b_cR^{ac}\\
&+(\frac{133}{2}\Pi_++\frac{47}{2}\Pi_-)K^{ac}K^b_c R_{anbn}+32\chi K^{ab}K^{cd}R_{acbd}+\frac{315}{2}K\Tr K^2S+(\frac{2041}{128}\Pi_++\frac{65}{128}\Pi_-)K^4\\
&+150 S\Tr K^3+(\frac{417}{32}\Pi_++\frac{141}{32}\Pi_-)K^2\Tr K^2+1080K^2S^2+360\Tr K^2 S^2\\
&+(\frac{375}{32}\Pi_+-\frac{777}{32}\Pi_-)(\Tr K^2)^2+\frac{885}{4}SK^3+(25\Pi_+-\frac{17}{2}\Pi_-)K\Tr K^3+2160KS^3\\
&+(\frac{231}{8}\Pi_++\frac{327}{8}\Pi_-)\Tr K^4-180E^2+180\chi E\chi E-120(\bar{\nabla} S)^2+720\chi(\bar{\nabla} S)^2\\
&-\frac{105}{4}\Omega_{ab}\Omega^{ab}+120\chi \Omega_{ab}\Omega^{ab}+\frac{105}{4}\chi\Omega_{ab}\chi\Omega^{ab}-45\Omega_{an}\Omega^a\,_n+180\chi \Omega_{an}\Omega^a\,_n-45\chi \Omega_{an}\chi\Omega^a\,_n\\
&-360\Omega^a\,_n \chi  \nb_a S+360\Omega^a\,_n  \nb_a S\chi-45\chi\nb_a\chi \Omega^a\,_n K-180\nb_a\chi\nb_b\chi\Omega^{ab}+30\chi\nb_a\chi\nb_b\chi\Omega^{ab}\\
&+90\chi\nb_a\chi\nabla_n\Omega^a\,_n+120\chi\nb^a\chi\nb^b\Omega_{ab}-180\chi\nb_a\chi\Omega_{bn}K^{ab}+300\nb_a\chi\nb^aE-180\nb_a\chi\nb^a\chi E\\
&-90\chi\nb_a\chi\nb^a\chi E+240\bar{\Box}\chi E-30\nb_a\chi\nb^a\chi R-60\nb_a\chi\nb_b\chi R^{ab}-30\nb_a\chi\nb_b\chi R_n\,^{ab}\,_n\\
&-\frac{675}{32}\nb_a\chi\nb^a\chi K^2-\frac{75}{4}\nb_a\chi\nb_b\chi K^{ac}K^b_c-\frac{195}{16}\nb_a\chi\nb^a\chi \Tr K^2-\frac{675}{8}\nb_a\chi\nb_b\chi KK^{ab}\\
&-330\nb_a\chi\nb^aSK-300\nb_a\chi\nb_bSK^{ab}+\frac{15}{4}\nb_a\chi\nb^a\chi\nb_b\chi\nb^b\chi+\frac{15}{8}\nb_a\chi\nb_b\chi\nb^a\chi\nb^b\chi-\frac{15}{4}(\bar{\Box}\chi)^2\\
&-\frac{105}{2}\nb_{a}\nb_b\chi\nb^a\nb^b\chi-15\nb_a\chi\nb^a\chi\bar{\Box}\chi-\frac{135}{2}\nb_a\chi\nb^a\bar{\Box}\chi\Big]\, .
\end{split}
\lb{a5}
\ee

\section{Algebraic equation for conformal charges in $d=5$}\label{appE}

\begin{equation*}
\scriptsize{
\begin{split}
&\Tr K^4\ : \ -6a+c_2+c_8 =261\, ,\\
&K\Tr K^3\ : \ 8a-c_2+2c_7-\frac{11}{6}c_8=-381\, ,\\
&(\Tr K^2)^2\ : \ 3a+c_1-2c_7=\frac{1251}{4}\, ,\\
&K^2\Tr K^2\ : \ -6a-\frac{1}{2}c_1+\frac{3}{8}c_2+\frac{47}{48}c_8=-66\, ,\\
&K^4\ : \ a+\frac{1}{16}c_1-\frac{3}{64}c_2-\frac{7}{48}c_8=\frac{267}{16}\, ,\\
&R_{acbd}^2\ : \ a+c_3 = 0 \, ,\\
&R_{ab}^2\ : \ -4a-\frac{16}{9}c_3+\frac{1}{9}c_4=0\, ,\\
&R^2\ : \ a+\frac{5}{18}c_3-\frac{1}{36}c_4=0\, ,\\
&R_{anbn}^2\ : \ -4a+c_4=0\, ,\\
&R^{ab}R_{anbn}\ : \ 8a+\frac{8}{3}c_3-\frac{2}{3}c_4=0\, ,\\
&R_{nn}^2\ : \ 4a+\frac{4}{9}c_3-\frac{4}{9}c_4=0\, ,\\
&RR_{nn}\ : \ -4a-\frac{8}{9}c_3+\frac{2}{9}c_4=0\, ,\\
&K^{ab}K^{cd}R_{acbd}\ : \ 4a+c_5-2c_7+\frac{2}{3}c_8 = 450\, ,\\
&K^a_cK^{bc}R_{ab}\ : \ 8a+\frac{2}{3}c_5-\frac{1}{3}c_6+2c_7-c_8 =-210\, ,\\
&KK^{ab}R_{ab}\ : \ -8a-\frac{5}{6}c_5+\frac{1}{6}c_6 =-60\, , \\
&\Tr K^2R\ : \ -2a-\frac{1}{12}c_5+\frac{1}{12}c_6+\frac{1}{3}c_8 =-60\, ,\\
&K^2R\ : \ 2a+\frac{1}{8}c_5-\frac{1}{24}c_6-\frac{5}{48}c_8 =30\, ,\\
&K^a_cK^{bc}R_{anbn}\ : \ -8a+c_6-2c_7+\frac{5}{3}c_8=810\, ,\\
&KK^{ab}R_{anbn}\ : \ 8a+\frac{1}{2}c_5-\frac{1}{2}c_6-\frac{1}{3}c_8 =-240\, ,\\
&\Tr K^2R_{nn}\ : \ 4a-\frac{1}{3}c_6-c_8=30\, ,\\
&K^2R_{nn}\ : \ -4a-\frac{1}{6}c_5+\frac{1}{6}c_6+\frac{11}{24}c_8 =-45\, ,\\
&\nb_cK_{ab}\nb^cK^{ab}\ : \ 2c_7-\frac{1}{3}c_8=-300\, ,\\
&\nb_aK^a_b\nb_cK^{bc}\ : \ -\frac{8}{3}c_7+\frac{8}{9}c_8=240\, ,\\
&\nb_aK^{ab}\nb_bK\ : \ \frac{4}{3}c_7-\frac{7}{9}c_8=0\, ,\\
&(\nb K)^2\ : \ -\frac{2}{3}c_7+\frac{25}{72}c_8=15\, ,\\
&K^{ab}\nabla_n R_{anbn}\ : \ -\frac{2}{3}c_8=240\, ,\\
&K\nabla_n R\ : \ \frac{1}{12}c_8=-30\, .
\end{split}}
\end{equation*}
These equations have a unique solution that determines the values for the conformal charges given in the main text.

\end{document}